\documentclass[12pt]{article}%
\usepackage[nosort]{cite}
\usepackage[usenames, dvipsnames]{xcolor}
\usepackage{graphicx}
\usepackage{multicol}
\usepackage{amsfonts}
\usepackage{amssymb}
\usepackage{amsmath}
\usepackage{heck}
\usepackage{braket}
\usepackage{afterpage}
\usepackage{setspace}
\usepackage{verbatim}
\usepackage{color}
\usepackage[normalem]{ulem}
\usepackage{longtable}
\usepackage{float}
\usepackage{subcaption}
\usepackage{epsfig}
\usepackage{enumerate}
\usepackage{epstopdf}
\usepackage[enableskew, vcentermath]{youngtab}
\usepackage{adjustbox}
\usepackage{multirow}
\usepackage{tikz}
\usepackage[margin=1in]{geometry}
\usepackage{titletoc}
\usepackage{hyperref}
\usepackage[percent]{overpic}
\usepackage{gensymb}
\usepackage{mathtools}%
\usepackage{tikz-cd}
\providecommand{\U}[1]{\protect\rule{.1in}{.1in}}
\pdfoutput=1
\newsavebox{\mysavebox}

\hypersetup{colorlinks,citecolor=black,filecolor=black,linkcolor=black,urlcolor=black}
\usetikzlibrary{decorations.markings}

\numberwithin{equation}{section}

\usetikzlibrary{chains}
\allowdisplaybreaks
\tikzset{node distance=2em, ch/.style={circle,draw,on chain,inner sep=2pt},chj/.style={ch,join},every path/.style={shorten >=4pt,shorten <=4pt},line width=1pt,baseline=-1ex}

\newcommand{\ba}{\begin{eqnarray}}
\newcommand{\ea}{\end{eqnarray}}

\newcommand{\be}{\begin{equation}}
\newcommand{\ee}{\end{equation}}

\tikzstyle{startstop} = [rectangle, rounded corners, minimum width=3cm, minimum height=1cm,text centered, draw=black, fill=blue!10]
\tikzstyle{startstop} = [rectangle, rounded corners, minimum width=3cm, minimum height=1cm,text centered, draw=black, fill=blue!10]
\tikzstyle{io} = [trapezium, trapezium left angle=70, trapezium right angle=110, minimum width=3cm, minimum height=1cm, text centered, draw=black, fill=blue!30]
\tikzstyle{process} = [rectangle, minimum width=3cm, minimum height=1cm, text centered, draw=black, fill=orange!30]
\tikzstyle{decision} = [diamond, minimum width=3cm, minimum height=1cm, text centered, draw=black, fill=green!30]
\tikzstyle{arrow} = [thick,->,>=stealth]
\tikzset{->-/.style={decoration={
  markings,
  mark=at position #1 with {\arrow[scale=2.4]{>}}},postaction={decorate}}}
\makeatletter \@addtoreset{equation}{section} \makeatother

\begin{document}

\rightline{IFT-UAM/CSIC-20-105 }

\date{July 2020}

\title{6D SCFTs, 4D SCFTs, \\[4mm] Conformal Matter, and Spin Chains}

\institution{UAM}{\centerline{${}^{1}$Instituto de Fisica Teorica UAM-CSIC, Cantoblanco, 28049 Madrid, Spain}}

\institution{PENN}{\centerline{${}^{2}$Department of Physics and Astronomy, University of Pennsylvania, Philadelphia, PA 19104, USA}}

\authors{Florent Baume\worksat{\UAM}\footnote{e-mail: {\tt florent.baume@uam.es}}, Jonathan J.\ Heckman\worksat{\PENN}\footnote{e-mail: {\tt jheckman@sas.upenn.edu}}, and Craig Lawrie\worksat{\PENN}\footnote{{e-mail: {\tt craig.lawrie1729@gmail.com}}}}

\abstract{Recent work has established a uniform characterization of
most 6D SCFTs in terms of generalized quivers
with conformal matter. Compactification of the partial tensor branch deformation of these theories
on a $T^2$ leads to 4D $\mathcal{N} = 2$ SCFTs which are also generalized quivers.
Taking products of bifundamental conformal matter
operators, we present evidence that there are large R-charge sectors of the
theory in which operator mixing is captured by a 1D\ spin chain Hamiltonian
with operator scaling dimensions controlled by a perturbation series in inverse powers of the
R-charge. We regulate the inherent divergences present in the 6D computations with the
associated 5D\ Kaluza--Klein theory. In the case of 6D SCFTs
obtained from M5-branes probing a $\mathbb{C}^{2}/\mathbb{Z}_{K}$
singularity, we show that there is a class of operators where the leading order mixing
effects are captured by the integrable Heisenberg $XXX_{s=1/2}$ spin chain with open boundary conditions,
and similar considerations hold for its $T^2$ reduction to a 4D $\mathcal{N}=2$ SCFT.
In the case of M5-branes probing more general D- and E-type
singularities where generalized quivers have conformal matter, we
argue that similar mixing effects are captured by an integrable $XXX_{s}$ spin chain
with $s>1/2$. We also briefly discuss some generalizations
to other operator sectors as well as little string theories.}

\maketitle

\setcounter{tocdepth}{2}

\tableofcontents

\enlargethispage{\baselineskip}

\newpage

\section{Introduction} \label{sec:INTRO}

One of the welcome surprises from string theory is the prediction of entirely
new classes of quantum field theories, such as interacting conformal fixed
points in six spacetime dimensions (see e.g. \cite{Witten:1995zh,
Strominger:1995ac, Seiberg:1996qx}). A remarkable feature of all higher-dimensional fixed points is that they are
``non-Lagrangian'' in the sense that they cannot be constructed from
perturbations of a Gaussian fixed point produced from free fields. By the same
token, this significantly complicates the study of such theories since many
textbook techniques based on perturbation theory are seemingly inapplicable.

In spite of these difficulties, the mere existence of higher-dimensional fixed
points provides a useful tool in the study of lower-dimensional systems. For
example, compactifications of 6D superconformal field theories (6D SCFTs)
produces a wealth of new sorts of lower-dimensional quantum field theories.
Additionally, dualities of known 4D quantum field theories can be understood
in terms of suitable compactifications of 6D SCFTs (see e.g.
\cite{Vafa:1997mh, Witten:1997sc, Gaiotto:2009we, Razamat:2019vfd}).  Clearly,
it would be desirable to better understand the structure of such systems, both
as a subject of interest in its own right, and also in terms of possible
lower-dimensional applications.

\begin{figure}[t!]
  \centering
  \includegraphics[scale=0.5]{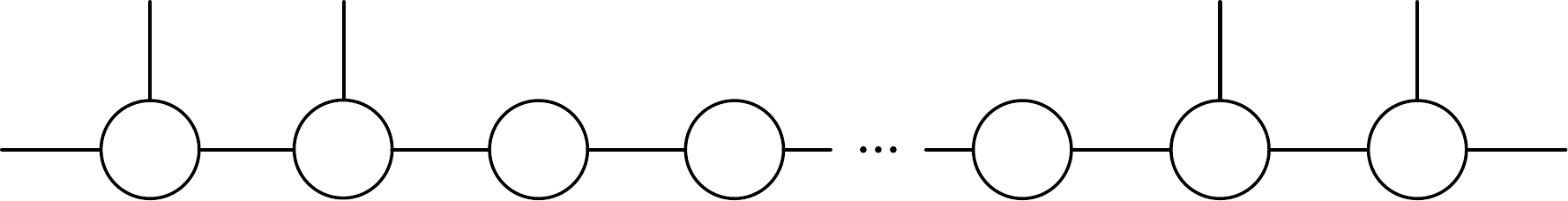}
  \caption{Depiction of the partial tensor branch of a generic 6D SCFT. These
  theories resemble generalized quiver gauge theories in which the links
  consist of conformal matter connecting gauge groups, as denoted by circles.
  Further decorations at the ends are possible.}
\label{fig:genquiv}
\end{figure}

One of the original ways to construct and study examples of such theories has
been through string compactification on singular geometries
\cite{Witten:1995zh}. Recent progress includes a classification of all
singular F-theory backgrounds which can generate a 6D SCFT
\cite{Heckman:2013pva, Heckman:2015bfa} (see also \cite{Bhardwaj:2015xxa, Bhardwaj:2019hhd} and
\cite{Heckman:2018jxk} for a review). A perhaps surprising outcome of this analysis is that on
a partially resolved phase of the singular geometry known as the partial tensor branch,
all known theories have a quiver-like structure which typically consists of a
single spine of ADE gauge group factors which are connected by 6D conformal
matter (see figure \ref{fig:genquiv} for a depiction). The geometric
realization provides direct access to the moduli space of these theories.

Complementary methods of study for 6D SCFTs include the use of the conformal
bootstrap \cite{Beem:2015aoa, Chang:2017xmr, Alday:2020lbp},
as well as the construction and study of holographic duals
(see e.g. \cite{Apruzzi:2013yva, Gaiotto:2014lca,
DelZotto:2014hpa, Filippas:2019puw, Bergman:2020bvi}).
Both have proven useful in extracting some information on
the operator content of 6D SCFTs, though it is fair to say that a more
complete understanding is still to be achieved. In particular,
extracting the explicit spectrum of operators and scaling dimensions in 6D SCFTs
has proven to be quite challenging.

Our aim in this paper will be to better understand the operator content of 6D
SCFTs, as well as their 4D descendants obtained from dimensional reduction. We
present evidence that in the limit where the length of a generalized quiver
becomes sufficiently long, there is a subsector of ``nearly-protected''
operators which have large R-charge $J$. In a sense we make precise, we find
that the scaling dimension for these operators can be organized as a perturbation series above a
bare scaling dimension $\Delta_0$:
\begin{equation}\label{pertJ}
\Delta = \Delta_{0} + \frac{\alpha}{J^2} + \mathcal{O}(J^{-3}),
\end{equation}
that is, we identify a perturbative expansion in large R-charge, and use it to
extract details of operator mixing in the 6D SCFT. This is very much in the
spirit of lower-dimensional examples where large R-charge limits were
fruitfully applied, as in reference \cite{Berenstein:2002jq}, as well as
\cite{Hellerman:2015nra}.

The operating assumption we make throughout this paper is that the 6D
conformal matter appearing as links in the generalized quiver description of
all 6D SCFTs can be used to define a class of operators in the 6D SCFT which trigger Higgs
branch deformations. Indeed, this picture was used in \cite{DelZotto:2014hpa,
Heckman:2014qba} to show that complex structure deformations of the F-theory
background can be interpreted as vacuum expectation values (vevs)
for operators in the accompanying SCFT.
For the most part, these rules are quite similar to Higgsing involving
weakly coupled hypermultiplets \cite{Gaiotto:2014lca, DelZotto:2014hpa,
Heckman:2014qba, Heckman:2016ssk, Heckman:2018pqx, Hassler:2019eso},
though it was also observed in \cite{Heckman:2014qba} that the scaling dimensions for these operators are
always significantly higher than that of a weakly coupled hypermultiplet.

By assumption, giving a vev to one such operator triggers a Higgs branch
deformation, and on the Higgs branch, we can study the resulting Nambu--Goldstone bosons. These bosons transform in a spin $s$ representation of $SU(2)_{\mathcal{R}}$, as dictated by the scaling dimension of the conformal matter operators. In a generalized quiver with $N$ gauge group factors $G_{i}$, and flavor symmetries $G_0$ and $G_{N+1}$ denoted via square brackets, which has the form:
\begin{equation}\label{eq:linearQuiver}
[G_0] - G_1 - \dots - G_{N} - [G_{N+1}] \,,
\end{equation}
we can, on the Higgs branch, visualize each link as a collection of Goldstone modes in a representation of $SU(2)_{\mathcal{R}}$ R-symmetry.
For the bifundamental between $G_i \times G_{i+1}$,
we will typically label these modes as $X^{(m_i)}_{i}$ for $-s \leq m_i \leq
s$, and for the highest, respectively lowest, weight we shall use the
simplified notation $X_i$, respectively $Y_i^{\dag}$.

Assuming the existence of these operators, we get a tremendous amount
of mileage in building gauge invariant combinations which survive as we move to the origin
of the tensor branch.
As an example, we can construct the gauge invariant composite bifundamental operator:
\begin{equation}\label{long}
	\mathcal{O}_{\mathrm{pure}} = \sqrt{\mathcal{Z}_N} X_{0} ... X_{N}\,,
\end{equation}
where the normalization factor, $\mathcal{Z}_N$, depends on the number of
fields and gauge groups, and is chosen such that the two-point function of
$\mathcal{O}_\text{pure}$ has coefficient one. Similar protected operators were
considered in \cite{Bergman:2020bvi}.

\begin{figure}[t!]
  \centering
  \includegraphics[scale=0.4]{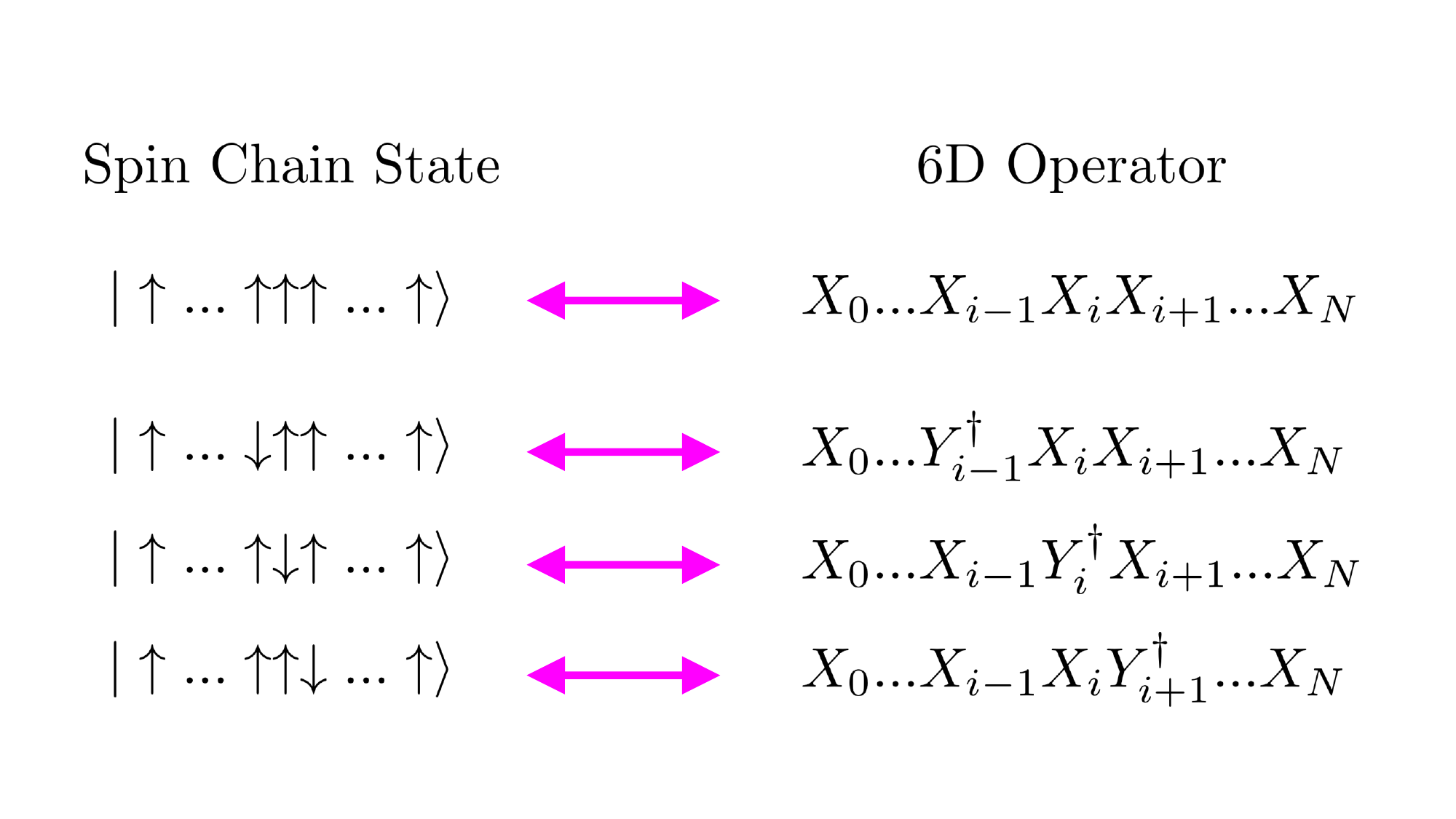}
  \caption{Depiction of the proposed correspondence between spin chain states and 6D operators. Here we consider the special case
of a 6D SCFT which has an A-type quiver gauge theory on its tensor branch, in which case the spin excitations are all spin $s = 1/2$ representations of the $SU(2)_{\mathcal{R}}$ R-symmetry group. Here, $X \oplus Y^{\dag}$ denotes the degrees of freedom of a bifundamental hypermultiplet in which $X$ denotes the spin up state and $Y^{\dag}$ denotes the spin down state. These operators are constructed on
the partial tensor branch of the 6D SCFT. Actual operators of the 6D SCFT are obtained by imposing a further decoupling constraint
which amounts to requiring zero total momentum for quasi-particle excitations.}
\label{fig:spinchain}
\end{figure}

Owing to the R-symmetry and flavor symmetry content of $\mathcal{O}_{\mathrm{pure}}$,
we expect it to have a protected scaling dimension proportional to $(N+1)$, the number of
generalized bifundamentals appearing in the product.
We can also consider descending to lower weight states for each $X_i$. Doing so we
get gauge invariant composite operators such as:
\begin{equation}\label{opopo}
\mathcal{O}_{m_0,...,m_N} = \sqrt{\mathcal{Z}_N} X^{(m_0)}_{0} ... X^{(m_N)}_{N},
\end{equation}
which has the structure of a 1D spin chain with each site a spin $s$ representation of $SU(2)_{\mathcal{R}}$.
While the highest weight state is $1/2$-BPS and protected, in 6D and for the analogous operators in 4D, we expect
there is operator mixing for other values of the $m_i$.
This leads to a correspondence between states of a spin chain and local operators:
\begin{equation}
\left|m_{0} ,..., m_{N}\right> \longleftrightarrow \mathcal{O}_{m_0,...,m_N}.
\end{equation}
See also figure \ref{fig:spinchain}.

We show that this operator mixing can be phrased in terms of a
spin chain with nearest neighbor hopping terms. A perturbative analysis on the tensor branch
of the 6D SCFT reveals that the two-point functions for ``neighboring'' impurity insertions are indeed non-zero, but that in the large $N$ limit,
the amount of such mixing is actually quite small. Indeed, in a diagonalized operator basis, we find that the eigenvalues
for such hopping terms are of order $g^2 / N^2$, where $g$ denotes a dimensionful coupling constant obtained from working on the
tensor branch.  One of the main observations that we will make in this paper is that a hopping term of this form will provide an indication that certain subsectors of the theory have operator mixing controlled by a 1D spin chain.

Of course, if our ultimate goal is to study operators at the conformal fixed point, we must find a way to return to strong coupling. To accomplish this, we consider the string theory background obtained from compactifying on a further circle. Retaining all of the Kaluza--Klein modes, we can treat this as a 5D ``Kaluza--Klein'' (5D KK theory) in which operator dimensions have their 5D values, but in which local operators are allowed to have support on all six spacetime dimensions. Using the embedding of this 5D KK theory in a string compactification, we can fix the value of the gauge coupling and evaluate the resulting hopping terms. This results in a matrix of anomalous dimensions, in accord with similar results obtained in the four-dimensional case where there are marginal coupling constants.

We perform this computation of operator mixing for a variety of 4D and 6D theories, beginning with the cases where we have the most
control, i.e. where $G_i = SU(K)$ for all $i$. As far as we are aware, the type of operator mixing we consider has not been studied previously even in the 4D case, the closest analog being the ``T-dual'' computations performed in references \cite{Beisert:2005he, Gadde:2010zi, Pomoni:2019oib} which also presented tantalizing hints of integrability in 4D $\mathcal{N} = 2$ SCFTs. With this in place, we then consider the case of a 6D SCFT with just $SU(K)$ gauge group factors, illustrating the close similarity with the 4D case. Applying our 5D KK regulator, we show that we again get a controlled perturbative expansion inversely in the R-charge of our operators.
In this case, the spin chain in question consists of spin $s = 1/2$ excitations and operator mixing is controlled by the
Heisenberg spin chain Hamiltonian \cite{Heisenberg:1928mqa}:
\begin{equation}
H_A = - \lambda_{A} \underset{i}{\sum} \left( 2 \overrightarrow{S}_{i} \cdot \overrightarrow{S}_{i+1} - \frac{1}{2} \right),
\end{equation}
where the constant $\lambda_{A}$ is computable both in 4D and 6D. The spectrum of energies in this
theory corresponds to the spectrum of anomalous dimensions for operators in this subsector. Importantly,
this Hamiltonian defines an integrable system and as such the quasi-particle spectrum is amenable
to methods such as the Bethe ansatz \cite{Bethe:1931} and its modern incarnations (see e.g. \cite{Faddeev:1996iy}),
and has figured prominently in the study of integrability in $\mathcal{N} = 4$
super Yang--Mills theory (see e.g. \cite{Minahan:2002ve} and
reference \cite{Beisert:2010jr} for an overview). So, we immediately gain a great deal of insight into the operator spectrum of
6D SCFTs. One can also consider generalizations of the A-type quivers in which the ranks of the gauge groups are not all constant.
This leads to a broader class of spin chain Hamiltonians,
and which in turn lead to modified dispersion relations for quasi-particle excitations.

Similar structure persists in the case of generalized quivers with D- and E-type gauge groups, though here, the spin excitations
are associated with conformal matter operators, and so we have a more general spin chain with spin $s > 1/2$ excitations. The important
point for us is that the holographic duals of all these cases are rather similar, being given by the
M-theory background $AdS_7 \times S^4 / \Gamma_{ADE}$
with $\Gamma_{ADE}$ a finite subgroup of $SU(2)$ (see e.g. \cite{DelZotto:2014hpa}).
This similarity provides a strong hint that the class of
excitations give in (\ref{opopo}) for the D- and E-series should also
be controlled by an integrable spin chain. Making the well-motivated assumption that integrability persists for
the D- and E-series, we also show how to extract the related spin chain
Hamiltonians for all the other cases. This is in turn controlled by
integrability of the $XXX_s$ spin chain, and the form of the Hamiltonian is
then:
\begin{equation}
H_{G} = - \lambda_{G} \underset{i}\sum{Q_{2s}(\overrightarrow{S}_{i} \cdot \overrightarrow{S}_{i+1})},
\end{equation}
where $Q_{2s}$ is a polynomial of degree $2s$ with relative coefficients fixed by the condition of integrability. In this case, our task
reduces to determining the constant $\lambda_{G}$, something we carry out for all of the related 4D and 6D SCFTs.

In all these cases, the spectrum of excitations is again controlled by a spin chain with open boundary conditions. We note that the case of periodic boundary conditions is also of interest and leads to a characterization of some operators in the little string theory (LST) obtained by gauging the diagonal subgroup of $G_0 \times G_{N+1}$ (see reference \cite{Bhardwaj:2015oru}). LSTs are especially intriguing because
even though they are inherently non-local (at high energies), they have a low energy effective field theory with operator content closely
related to their 6D SCFT counterparts.

Though we primarily focus on the operators of line (\ref{opopo}), the topology
of these generalized quivers also permits us to construct related spin chains.
As an example, we can consider operators such as:
\begin{equation}
	\mathcal{B}_{i} = \sqrt{\mathcal{Z}_{\mathcal{B}_i}} X_0 ... X_i (Y_i X_i) X_{i+1} ... X_N,
\end{equation}
and track the movement of the $(Y_i X_i)$ insertion. We can also construct
closed loops in a generalized quiver such as:
\begin{equation}\label{loop}
	\mathcal{C}_{i,i+L} =  \sqrt{\mathcal{Z}_{\mathcal{C}_{i,i+L}}} \mathrm{Tr}(X_{i} ... X_{i + L} Y_{i+L} ... Y_{i})
\end{equation}
in the obvious notation. The level of protection from operator mixing is lower in these cases,
since there are transitions to multi-trace operators. Such transitions can be suppressed if we also
assume that the rank of the gauge groups in the generalized quiver are sufficiently large so that only
planar diagrams contribute. Provided the R-charge (i.e. the length of the spin
chain) is large enough, we again find a perturbative expansion in inverse powers of the R-charge.
This leads to a quite similar analysis for impurity insertions and operator mixing,
but with different boundary conditions for the associated
spin chain problem.

A pleasant feature of the $\mathcal{C}$ spin chain
operators is that in the large $N$ limit, perturbations can also be detected in
the holographic dual theories, provided we also take $L \sim N^{1/3}$. Indeed,
this leads to the pp-wave limit of the geometry $AdS_{7} \times S^{4} /
\Gamma_{ADE}$, the same sort studied in \cite{Berenstein:2002jq, Berenstein:2002zw}.
In the holographic dual with orbifold fixed points of $S^{4} / \Gamma_{ADE}$ at the
north and south pole, the original operators of interest correspond to
gravitons with large momenta orbiting along a fixed latitude, the precise
location of which depends on the values of $i$ and $i+L$ in equation \eqref{loop}.
Again, we note that unless we also assume that the ranks of the flavor groups scale to large
size so as to remain in the planar limit, there is significant mixing with multi-trace operators.

The rest of this paper is organized as follows. We begin in section
\ref{sec:GENQUIV} by reviewing the generalized quiver picture of 6D SCFTs, and
in particular present our main hypotheses and assumptions on the properties of
6D conformal matter. With this in place, we turn to some examples of
quivers with A-type gauge groups, considering the case of 4D SCFTs
in section \ref{sec:4DATYPE} and 6D SCFTs in section \ref{sec:6DATYPE}. In
particular, we establish the existence of a nearly-protected sector of
operators with mixing controlled by a matrix of anomalous dimensions which
resembles ``hopping terms'' in a 1D spin chain. Following this, we turn in
section \ref{sec:CONFTYPE} to a further generalization of these considerations
to generalized quivers with D- and E-type quivers, both for 4D and 6D SCFTs.
We present our conclusions in section \ref{sec:CONC}. Some
additional technical details are presented in the Appendices.

\section{6D SCFTs as Generalized Quivers \label{sec:GENQUIV}}

In this section we briefly review some aspects of 6D SCFTs, in particular the fact that on a partial tensor branch they all resemble generalized quivers. Our aim will be to exploit this structure to extract additional details on the operator content of these fixed points. With this in mind, we first briefly review the construction of these theories, both in F-theory and M-theory. We then turn to an analysis of 6D conformal matter, and in particular the expectation that there are specific operators which can be used to build large composite operators.

\subsection{Top Down Construction of 6D SCFTs}

To begin, let us briefly review the top down construction of 6D SCFTs. The starting point for all known constructions
involves F-theory on a non-compact elliptically fibered Calabi--Yau threefold $X \rightarrow B$.\footnote{We note that
even for theories with a frozen phase \cite{Witten:1997bs, Tachikawa:2015wka, Bhardwaj:2018jgp},
there is a geometric avatar \cite{Bhardwaj:2015oru, Bhardwaj:2018jgp}.} A
6D SCFT is obtained by seeking out configurations of curves which can all simultaneously collapse to zero size inside the base $B$. The general feature found in reference \cite{Heckman:2013pva, Heckman:2015bfa} is that such contractible configurations of curves in the base all have a rather uniform structure, approximately assembling into a single line of collapsing curves with a small amount of decoration on the left and right sides of such a configuration. In fact, in subsequent work it was realized that all of these examples descend from a handful of ``progenitor theories'' under a process of fission and fusion \cite{Heckman:2018pqx}. These theories are precisely the ones which can be realized from M5-branes probing an ADE singularity wrapped by the M9-brane wall of heterotic M-theory. Our primary interest in this paper will be on the closely related examples obtained by a single tensor branch deformation, where we pull the M5-branes off the $E_8$ nine-brane wall, so that they just probe the space $\mathbb{R}_{\bot} \times \mathbb{C}^{2} / \Gamma_{ADE}$.

In the M-theory realization, we can think of the ADE singularity as generating
a 7D super Yang--Mills (SYM) theory coupled to a gravitino multiplet (see \cite{DelZotto:2014hpa, Heckman:2014qba, Heckman:2017uxe}). Introducing $N+1$ probe M5-branes realizes a domain wall with localized states trapped on the wall. This also makes it clear that we get a $G_L \times G_R$ flavor symmetry associated with the ADE singularity. Separating the M5-branes in the $\mathbb{R}_{\bot}$ direction corresponds to moving onto the ``partial tensor branch.'' In this picture, each finite length segment produces a compactification of 7D SYM which preserves 6D $\mathcal{N} = (1,0)$ supersymmetry on the wall.

Similar considerations hold in the F-theory realization of these theories. In this case, each finite interval is instead associated with a curve of self-intersection $-2$ which is wrapped by a seven-brane with gauge group $G_{ADE}$, and the half-lines to the left and the right correspond to non-compact curves. At each collision of seven-branes we have localized matter. In all cases other than A-type seven-branes, further blowups in the base are required to reach a smooth F-theory model. In the M-theory realization of these theories, this corresponds to a further fractionation of the M5-branes \cite{DelZotto:2014hpa}. The ``6D conformal matter'' theories with the same flavor symmetry factor $G-G$ are then given by (see also \cite{Bershadsky:1996nu, Aspinwall:1997ye, Morrison:2012np}):
\begin{equation}\label{eqn:bbb}
\begin{aligned}
[D_K] - [D_K] & \equiv [D_K], \overset{\mathfrak{sp}_{K-4}}{1} , [D_K]\cr
[E_6] - [E_6] & \equiv [E_6], 1, \overset{\mathfrak{su}_3}{3}, 1, [E_6] \cr
[E_7] - [E_7] & \equiv [E_7], 1, \overset{\mathfrak{su}_2}{2},\overset{\mathfrak{so}_{7}}{3},\overset{\mathfrak{su}_2}{2},1, [E_7] \cr
[E_8] - [E_8] & \equiv [E_8], 1, 2, \overset{\mathfrak{sp}_1}{2},\overset{\mathfrak{g}_2}{3},1,\overset{\mathfrak{f}_4}{5},1,\overset{\mathfrak{g}_2}{3},
\overset{\mathfrak{sp}_1}{2},2, 1, [E_8].
\end{aligned}
\end{equation}
Here, each number $n$ denotes a smooth rational curve of self-intersection $-n$ in the base $B$. In the 6D theory it is associated with a tensor multiplet of that charge. Each superscript over a curve denotes a gauge algebra. For brevity we have suppressed the bifundamental matter which appears from additional collisions of seven-branes, and as required by anomaly cancellation considerations.

To build a generalized quiver theory, we consider pairs of conformal matter theories $G - G$ and $G - G$ and gauge a diagonal subgroup, reaching the theory $G - (G) - G$. This gauging procedure must be accompanied by an additional tensor multiplet to cancel gauge anomalies. This process was referred to as a fusion operation in reference \cite{Heckman:2018pqx}. Doing so, we reach a generalized quiver of the form:
\begin{equation}\label{eqn:qqq}
[G_0] - G_1 - ... - G_N - [G_{N+1}],
\end{equation}
corresponding to $N + 1$ M5-branes probing the singularity
$\mathbb{C}^2 / \Gamma_{ADE}$ of the same ADE type as $G$.

We will now briefly explain the distinction between the full and the partial tensor branch of the quivers (\ref{eqn:qqq}). The full tensor branch is given by the geometric configuration where all of the curves in the F-theory base have non-zero volume: between each gauge group $G_i = G_{i+1}$ there exist all the conformal matter curves appearing in (\ref{eqn:bbb}). This full tensor branch description is the generic description of the theory at a general point of the tensor branch. On the full tensor branch the 6D theory has no tensionless string-like degrees of freedom. The partial tensor branch occurs at the higher codimension point on the tensor branch where the volumes of all the conformal matter curves are taken to zero, but the volumes of the curves supporting the gauge groups $G$ remain finite.

An important feature of the partial tensor branch theory is that
further compactification on a $T^2$ results in a 4D $\mathcal{N} = 2 $ SCFT
\cite{Ohmori:2015pua, Ohmori:2015pia} (see also \cite{Apruzzi:2018oge}).\footnote{In general, the compactification leads to a 4D SCFT coupled to additional vector multiplets, and decoupling these vector multiplets leads to the SCFT of interest here.} One piece of evidence for this is
obtained by evaluating the contribution of the 4D conformal matter to the beta function of a gauge group $G$. In
conventions where an $\mathcal{N} = 2$ vector multiplet has beta function coefficient $b_{\mathrm{vec}}(G) = 2h_{G}^{\vee}$, each conformal
matter link contributes as $b_{\mathrm{matt}}(G,G) = - h_{G}^{\vee}$, where $h_{G}^{\vee}$ denotes the dual Coxeter number of the gauge group.
This illustrates that although we are on the tensor branch, there is still a notion of conformality which survives to lower dimensions. An additional remark is that if we had moved to the full 6D tensor branch and \textit{then} compactified, we would have reached a 4D theory which is not conformal.

\subsection{Conformal Matter}

The presentation in terms of conformal matter is more than just suggestive
pictorially. For many purposes, the degrees of freedom localized at a link
behave like matter fields. This point of view was developed in
\cite{DelZotto:2014hpa, Heckman:2014qba} where it was noted that there is a
class of complex structure deformations in the associated Calabi--Yau threefold
which directly match to Higgs branch deformations of the 6D SCFT. The picture
of Higgsing in 6D SCFTs in terms of nilpotent orbits and the corresponding
match to vevs of generalized matter provides further support for this general
physical picture. With this in mind, our aim here will be to collect some
useful aspects of $(G,G)$ conformal matter for $G$ an ADE group.

As a preliminary comment, we note that in F-theory, each of these theories can
be realized as the collision of two seven-branes with gauge group $G$ which
collide over a common 6D spacetime. Locally, each of these seven-branes can be
modelled as an ADE singularity, so for our present purposes we can dispense
with the requirement of an elliptic fibration. The local structure of the
different conformal matter theories is then given by:
\begin{align}
	(A_{K-1},A_{K-1})  & :\,y^{2}=x^{2}+(uv)^{K},\\
	(D_{K},D_{K})      & :\,y^{2}=(uv)x^{2}+\left(  uv\right)  ^{K-1},\\
	(E_{6},E_{6})      & :\,y^{2}=x^{3}+(uv)^{4},\\
	(E_{7},E_{7})      & :\,y^{2}=x^{3}+(uv)^{3}x,\\
	(E_{8},E_{8})      & :\,y^{2}=x^{3}+(uv)^{5}
\end{align}
where $u$ and $v$ are local coordinates of the base. A natural deformation of this geometry is given by brane recombination of
two distinct stacks of seven-branes. In the local singularity, this amounts to a smoothing deformation of the form:
\begin{equation}
uv \mapsto uv - r.
\end{equation}
Though not originally stated in these terms, in reference \cite{Heckman:2014qba} the scaling dimension of this recombination operator was determined in the theory obtained by compactifying the 6D SCFT on a further $S^1$. Strictly speaking, this computation was performed in a 5D Kaluza--Klein (KK) theory, in which a free scalar would have scaling dimension $\Delta_{\mathrm{KK}} = 3/2$ rather than the 6D free hypermultiplet value of $\Delta_{\mathrm{hyper}} = 2$.\footnote{This point has been taken into account in a revised version of reference \cite{Heckman:2014qba}.} Taking this subtlety into account, we obtain a table of scaling dimensions for the recombination operators in the case of $N + 1$ M5-branes probing an ADE singularity:
\begin{equation} \label{taboom}
\begin{tabular}
[c]{|c|c|c|c|c|c|}\hline
& $(A_{K-1},A_{K-1})$ & $(D_{K},D_{K})$ & $(E_{6},E_{6})$ & $(E_{7},E_{7})$ &
$(E_{8},E_{8})$\\\hline
dim $r$ & $4(N+1)$ & $8(N+1)$ & $12(N+1)$ & $16(N+1)$ & $24(N+1)$\\\hline
\end{tabular}
\ \ .
\end{equation}
We note that in the case of the A-type singularity, the first non-trivial
fixed point arises at $N = 1$ (two M5-branes), as the $N=0$ case (one M5-brane)
is simply a free bifundamental hypermultiplet.

Now, at least in the case of the A-type conformal matter, we observe that a weakly coupled hypermultiplet $X \oplus Y^{\dag}$ in the bifundamental representation has scaling dimension $\Delta = 2$. From this, we conclude that at least in the case of a single M5-brane, the recombination
operator is associated with the vev of the combination $\mathrm{Tr}(XY)$. More generally, we can consider a ``classical quiver'' with $N$ such gauge group factors. In this case, we can construct the related operator
\begin{equation}
(A_{K-1} , A_{K-1}) ~ \mathrm{recombination \, operator:}\qquad
 r \sim \mathrm{Tr}(X_0 ... X_N Y_N ... Y_0) \,,
\end{equation}
in the obvious notation, and this has the expected scaling dimension
for the recombination operator. From this, we can already identify
a natural gauge invariant bifundamental operator:
\begin{equation}
    \mathcal{O}_{\mathrm{pure}} = \sqrt{\mathcal{Z}_N} X_0 ... X_N,
\end{equation}
which has scaling dimension $2(N+1)$. This operator is the highest weight scalar field inside of a $\mathcal{D}$-type superconformal multiplet. A short review of the 6D superconformal multiplets is given in Appendix \ref{app:superRepresentations}.
Similar protected operators
were considered in \cite{Bergman:2020bvi}. Let us note that even in the A-type case, we
are performing our analysis on the tensor branch, and one could of course
dispute whether this sort of operator survives at the conformal fixed point. In
the present case, however, the high amount of (super)symmetry, along with
the direct match to geometry provides good evidence that this assignment is
correct at the conformal fixed point as well.  In terms of the data directly
visible in the 6D SCFT, the operator $\mathcal{O}_{\mathrm{pure}}$ is in the
bifundamental representation of the symmetry group $G_0 \times G_{N+1}$.
Additionally, we note that since each $X_i \oplus Y_{i}^{\dag}$ hypermultiplet transforms in the
spin $1/2$ representation of $SU(2)_{\mathcal{R}}$ R-symmetry, the composite
formed from $N+1$ such operators transforms in an irreducible representation of
the tensor product:
\begin{equation}
    (1/2)^{\otimes(N+1)} = \frac{N+1}{2} \oplus ...
\end{equation}
namely the operator $\mathcal{O}_{\mathrm{pure}}$ is the highest weight state of a
spin $(N+1)/2$ representation of $SU(2)_{\mathcal{R}}$ R-symmetry.

A deceptively similar analysis also works in the case of generalized quivers
with $SO$ gauge groups provided we take $SO_{2K}$ with $K \geq 5$. Indeed, in
this case we can move onto the full tensor branch:
\begin{equation}
    [SO_{2K}] , \overset{\mathfrak{sp}_{K-4}}{1}, SO_{2K} - ... - \overset{\mathfrak{sp}_{K-4}}{1}, [SO_{2K}].
\end{equation}
Between each $SO / Sp$ factor there is a half hypermultiplet in the
bifundamental representation, so forming suitable composite operators we again
recover the scaling dimension of the recombination operator presented in
\eqref{taboom}. That being said, there are some indications
from the study of Higgs branch flows that performing this further tensor branch
deformation is actually inappropriate. One reason is that these weakly coupled
bifundamentals do not by themselves account for the full space of possible
Higgs branch deformations \cite{Heckman:2016ssk} (see also
\cite{Bourget:2019aer}). Another issue is that upon further compactification on
a $T^2$, this would not generate a 4D $\mathcal{N} = 2$ SCFT, indicating that
``too many'' degrees of freedom have been removed in this process. Finally, we
face the awkward fact that for $K = 4$, there are no gauge group factors at all
on the $(-1)$-curves!

To rectify this and to also give a uniform treatment of all the conformal
matter cases, we shall instead proceed differently. First, we give a heuristic
argument explaining the appearance of the precise scaling dimensions for the
recombination operators. Given a gauged node with gauge group $G_i$ situated
in a generalized quiver as $G_{i-1} - G_i - G_{i+1}$, we can consider decompactifying the neighboring
factors so that the left and right neighboring factors become flavor symmetries. Performing blowdowns and smoothings of
the conformal matter links, we can consider deformations which break these flavor symmetry factors but leave intact
the gauge group $G_i$. Doing so, we get the
following pattern of geometries:
\begin{align}
[D_K] - D_K - [D_K]\quad\longrightarrow &\quad\overset{\mathfrak{sp}_{K-4}}{1}, \overset{\mathfrak{so}_{2K}}{4}, \overset{\mathfrak{sp}_{K-4}}{1} \\
[E_6] - E_6 - [E_6]\quad\longrightarrow &\quad2,1, \overset{\mathfrak{e}_{6}}{6}, 1,2 \\
[E_7] - E_7 - [E_7]\quad\longrightarrow &\quad2,2,1, \overset{\mathfrak{e}_{7}}{8}, 1,2,2 \\
[E_8] - E_8 - [E_8]\quad\longrightarrow &\quad2,2,2,2,1, \overset{\mathfrak{e}_{8}}{12}, 1,2,2,2,2.
\end{align}
Now, a tail of $1,2,...,2$ with $Q$ curves defines the tensor branch of the
rank $Q$ E-string theory with flavor symmetry $E_8$ from the M9-brane of
heterotic M-theory. What we are doing is taking a pair of such theories and
then gauging a diagonal subgroup of this $E_8$. When the tails are of the form $\overset{\mathfrak{sp}_{K-4}}{1}$, we are taking a pair of minimal $(D_K, D_K)$ conformal matter theories and gauging a common $SO(2K)$ inside of the $SO(4K)$ flavor symmetries. Suppose we now compactify on an
$S^1$. The resulting KK theory for the rank $Q$ E-string can be viewed as an
$\mathfrak{sp}_Q$ gauge theory coupled to $N_f = 8$ flavors in the fundamental
representation (see e.g. \cite{Seiberg:1996bd, Douglas:1996xp,
Morrison:1996xf}). Additionally, there is a hypermultiplet in the
two-index anti-symmetric representation of $\mathfrak{sp}_Q$, which we
denote as $A$. At the point of strong coupling the $SO(16)$ flavor
symmetry enhances to the affine $\widehat{E}_{8}$ symmetry (since we
are dealing with a 5D KK theory). In this $\mathfrak{sp}_{Q}$ gauge
theory there is a remnant of the recombination operator. Letting $H
\oplus \widetilde{H}^{\dag}$ denote the bifundamental hypermultiplet,
this is schematically of the form:
\begin{equation}
r_{KK} \sim (H \times A^{Q-1} \times \widetilde{H})_L \times (H \times A^{Q-1} \times \widetilde{H})_R,
\end{equation}
where here, we have included the operators of the two E-string theories to the
left and right. Note that each factor forms a gauge invariant operator of the
corresponding $\mathfrak{sp}_{Q}$ gauge theory.  Using the free field values of
these 5D KK modes, we get $\dim r_{KK} = 2 \times 3/2 \times (Q + 1)$ and so in
the lift to 6D, we get $\dim r = 2 \times 2 \times (Q + 1)$. We observe that for D-type -- where a similar argument can be made starting from minimal $(D_K, D_K)$ conformal matter instead of the E-string --
and $E_6,E_7$, and $E_8$ conformal matter, the respective values of $Q$ are
$1,2,3,5$, so we indeed recover the expected scaling dimensions for the
recombination operators, as given in reference \cite{Heckman:2014qba}.

Encouraged by this match, we shall therefore indeed assume the existence of a
conformal matter operator which can attain a vev, and in so doing initiates a
flavor symmetry breaking pattern. Now, because of the rather high dimension of
the associated operator, it is no longer appropriate to view the associated
operators as filling out an $SU(2)_{\mathcal{R}}$ R-symmetry doublet. Observe,
however, that once we trigger a vev for these fields, we can model the effects
of flavor symmetry breaking in terms of Nambu--Goldstone modes in the coset
space $(G_L \times G_R )/ G_{\mathrm{diag}} $ (in the case of diagonal
breaking). So, in spite of the fact that we are dealing with exotic matter, on
the Higgs branch of the theory we can still model the effects in terms of
perturbations of free fields. These modes also transform in a spin $s$
representation of the $SU(2)_{\mathcal{R}}$ R-symmetry where the spin
assignments are:
\begin{equation}%
\begin{tabular}
[c]{|c|c|c|c|c|c|}\hline
& $(A_{K-1},A_{K-1})$ & $(D_{K},D_{K})$ & $E_{6}$ & $E_{7}$ & $E_{8}$\\\hline
$s$ & $1/2$ & $1$ & $3/2$ & $2$ & $3$\\\hline
\end{tabular}
\ \ \ .
\end{equation}
We shall label each $X_{i}$ as $X_{i}^{(m_{i})}$
where $-s\leq m_{i}\leq s$ denoting the specific spin. Note that in
this notation, the hypermultiplet doublet $X_{i}\oplus Y_{i}^{\dag}$ would be written
as $X_{i}^{(+1/2)}\oplus X_{i}^{(-1/2)}$.

Weakly gauging the flavor symmetry of conformal matter, we can ask how these Goldstone modes now couple
to the corresponding vector multiplet. To leading order, we expect a term which relates the triplet of D-terms
to expressions which are quadratic in the Goldstone modes:
\begin{equation}
D_{i,a}^{R}=\frac{1}{s}\times\left( \mathrm{Tr}_{i+1} ( X_{i}^{\dag(m_{i})}S_{R}%
^{(m_{i},n_{j})}T_{i,a}X_{i}^{(n_{j})}) - \mathrm{Tr}_{i-1} (X_{i-1}^{(m_{i})}S_{R}%
^{(m_{i},n_{j})}T_{i,a}X_{i-1}^{\dag(n_{j})} ) \right)  + \cdots \,,
\end{equation}
with $R=1,2,3$ an $SU(2)_{\mathcal{R}}$ R-symmetry triplet index, and $S_{R}^{n,m}$ denotes a matrix entry
of a spin $s$ symmetry generator. Here, we have also
included the contributions from the Lie algebra generators. In the above, the
specific normalization has been chosen so that the highest weight state of the spin $s$ representation
couples with unit strength to the vector multiplet. Additionally, the
appearance of the \textquotedblleft$...$\textquotedblright\ indicates that
at least for $s > 1/2$, we expect higher order terms. In stringy terms,
we expect such corrections to be present because the
M5-branes fractionate at D- and E-type singularities, and this
fractionation means that degrees of freedom on an M5-brane can be viewed as composites from
these fractionalized pieces. Note that in the special case of $s = 1/2$ no such fractionation occurs,
and this is in accord with just taking
the minimal coupling between a 6D hypermultiplet and vector multiplet.

\subsection{Decoupling Limit \label{ssec:DECOUP}}

The main idea we will be developing in this paper is that there is a class of
operators of a 6D SCFT which can be realized by building gauge invariant operators on the
partial tensor branch of a 6D\ SCFT. In a suitable decoupling limit we expect some of these operators
to also be present at the conformal fixed point.

\begin{figure}[t!]
\begin{center}
 \includegraphics[scale=0.45]{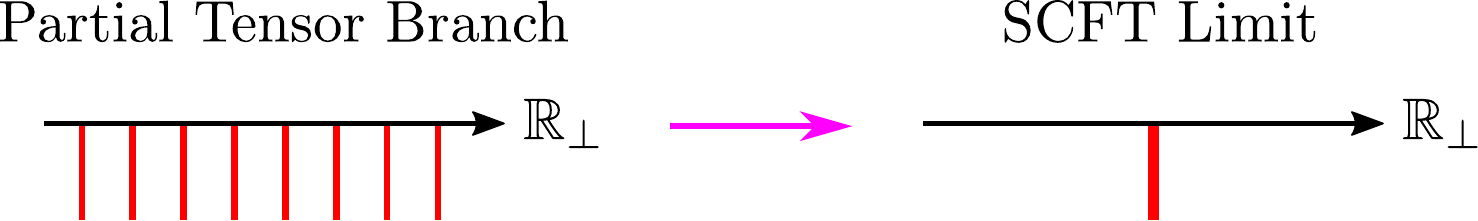}
\caption{Depiction of the partial tensor branch of M5-branes (red vertical lines) filling $\mathbb{R}^{5,1}$
and probing the transverse geometry $\mathbb{R}_{\bot} \times \mathbb{C}^{2} / \Gamma_{ADE}$.
Separating the M5-branes along the $\mathbb{R}_{\bot}$ direction generates a 1D lattice. The conformal
fixed point corresponds to the limit where all M5-branes coincide. Localized fluctuations on the 6D domain wall
are those which are annihilated by the translation operator $P_{\bot}$ and its discretized analog
on the partial tensor branch.}
\label{fig:M5lattice}
\end{center}
\end{figure}

In terms of the M-theory realization of the 6D\ SCFT via M5-branes
probing the geometry $\mathbb{R}_{\bot}\times\mathbb{C}^{2}/\Gamma_{ADE}$,
the partial tensor branch is reached by keeping the M5-branes at the orbifold
singularity and separating them along the $\mathbb{R}_{\bot}$ direction. Doing
so, we see that each conformal matter sector is associated with an edge mode
localized on an M5-brane. We observe here that in addition to fluctuations
along a given M5-brane, there can also be fluctuations of states in the
$\mathbb{R}_{\bot}$ direction (the bulk). We can view the locations of
M5-branes with equal relative separations in the $\mathbb{R}_{\bot}$ direction
as specifying a 1D\ lattice. See figure \ref{fig:M5lattice} for a depiction of the
1D lattice realized by M5-branes on the partial tensor branch. Letting
$\ell_{\ast}$ denote this lattice spacing, we can see that momenta in the
$P_{\bot}$ direction will be quantized in units of $ \ell_{\ast} /(N+1)$.
The discretized momentum operator of the lattice acts
on an operator at the $j^{th}$ lattice site as:
\begin{equation}
e^{-iP_{\bot}}X_{j}e^{iP_{\bot}}=X_{j+1}.
\end{equation}
The symmetry is broken by boundary effects of $\mathbb{R}_{\bot}$, but it is
retained in the closely related situation where we instead compactify on an
$S^{1}$ (which would lead to a little string theory).
The decoupling limit used to reach an
SCFT corresponds to sending $\ell_{\ast}\rightarrow 0$. So, any excitation with
finite lattice momentum becomes a highly excited state in this limit. Of
course, this also means that the purely localized states associated with the
CFT are those for which the total momentum in the $\mathbb{R}_{\bot}$
direction is exactly zero.

The states of the CFT\ are those
which survive the decoupling limit where we also send all M5-branes
on top of each other. Indeed, any excitation with non-zero $P_{\bot}$
momentum becomes a highly excited state in this limit.
This in turn enforces the condition that any edge mode decoupled
from the bulk satisfies the additional condition:%
\begin{equation}
P_{\bot}\left\vert \mathcal{O}_{SCFT}\right\rangle =0,\label{Pbot}%
\end{equation}
for any state of the SCFT.

How to enforce this condition in practice? We follow a pragmatic approach
where at first, we allow all possible momentum excitations along our
1D\ lattice. For example, for operators such as:%
\begin{equation}
\mathcal{O}_{m_{0}...m_{N}}= \sqrt{\mathcal{Z}_N} X_{0}^{(m_{0})}...X_{N}^{(m_{N})},
\end{equation}
we can visualize the operators on the tensor branch as a specific
configuration of quantum spins in a 1D lattice. As has been appreciated for
some time in such 1D\ systems, there are quasi-particle excitations which can
be constructed out of these excitations known as magnon excitations. In fact,
we will develop further this quasi-particle
excitation picture. The important point for us is that these quasi-particles
carry a well-defined lattice momentum, and so we can impose the condition that
any quasi-particle excitations need to have zero total momentum. For $I$ such impurities,
the condition is then:
\begin{equation}
\text{6D\ Decoupling Constraint:} \,\,\, p_1 + ... + p_I = 0,
\end{equation}
where each $p_i$ denotes the momentum of a quasi-particle excitation.

To illustrate, consider a ground state of a ferromagnetic spin chain as
specified by:%
\begin{equation}
    \mathcal{O}_{s...s}= \sqrt{\mathcal{Z}_N} X_{0}^{(s)}...X_{N}^{(s)},
\end{equation}
which has all spins pointing up. There is no momentum in this excitation.

In the single impurity sector, we can consider operators constructed such as those
obtained by flipping one spin at the  $j^{\text{th}}$ lattice site:%
\begin{equation}
    \mathcal{O}_{j}= \sqrt{\mathcal{Z}_N} X_{0}^{(s)}...X_{j}^{(s-1)}...X_{N}^{(s)}.
\end{equation}
We emphasize that these are operators constructed on the tensor branch, and
there is no a priori guarantee that they will all survive in the decoupling
limit. Indeed, in this sector, there is a single $P_{\bot} = 0$ operator:
\begin{equation}
    \mathcal{O}_{\text{zero}}^{\text{1-imp}}= \frac{1}{\sqrt{N+1}} \left( \mathcal{O}_{0}+...+\mathcal{O}_{N} \right),
\end{equation}
which is just the descendant of $\mathcal{O}_{s...s}$ under
$SU(2)_{\mathcal{R}}$ R-symmetry. All other linear combinations in the single
impurity sector have non-zero momentum, and as such do not survive in the
decoupling limit.  Similar considerations hold in the presence of additional
impurities. In this case, it is convenient to state the ``zero-momentum
condition'' in terms of the quasi-particle excitations associated with the
algebraic Bethe ansatz.

We will also be interested in the quite similar class of 4D generalized
quivers obtained from dimensional reduction of the partial tensor branch
theory on a further $T^{2}$. In this setting, we can again ask whether a zero
momentum condition needs to be enforced here as well. In this case, there is
no need to do so, because even when we keep the M5-branes at finite separation
the resulting 4D\ system still realizes an SCFT. So, while we need to enforce
equation (\ref{Pbot}) in 6D\ SCFTs, in 4D\ SCFTs with the same quiver
structure, there is no such constraint.

\section{4D\ $\mathcal{N}=2$ SCFTs with Classical Matter \label{sec:4DATYPE}}

Having introduced the main features of generalized quivers in 6D SCFTs, our aim will
now be to understand operator mixing in some specific operator subsectors.

As a warmup exercise, and as a subject of interest in its own right, in this
section we consider 4D\ $\mathcal{N}=2$ SCFTs\ with A-type gauge groups
arranged along a linear quiver. We show that, much as in \cite{Berenstein:2002jq}
(see also \cite{Berenstein:2002zw, Berenstein:2005vf}),
there are subsectors of operators which exhibit operator mixing as controlled by a
1D\ spin chain, even when we pass to very large gauge coupling $g$. We
accomplish this by constructing an alternative perturbative expansion
in the large R-charge, $J$ of our operators. This expansion is valid in the regime where $g^2/J^2$
is small. It is important to note that the results of \cite{Berenstein:2002jq} do not carry over to the 4D $\mathcal{N}=2$ setup in general \cite{Nishioka:2008gz,Gromov:2014eha,Mitev:2015oty}. In our case, it is vital that the R-charge, $J$, corresponding to the length of the quiver, is large, and this adds a significant additional suppression beyond that of \cite{Berenstein:2002jq}.

As far as we are aware, the specific class of operators we consider has not
been previously studied, the closest analog being the analysis of \cite{Beisert:2005he} as well as
\cite{Gadde:2010zi} which feature a T-dual quiver gauge theory. We will comment on the relation
to the T-dual setup later.

Consider then the specific 4D $\mathcal{N}=2$ quiver gauge theory:%
\begin{equation}
\lbrack G_{0}]-G_{1}-...-G_{N}-[G_{N+1}],
\end{equation}
where here, each gauge group factor corresponds to $G_{i}=SU(K)$. In this
case, we have bifundamental hypermultiplets $X_{i}\oplus Y_{i}^{\dag}$ between
each link. There is a $1/2$-BPS\ operator of dimension $\Delta=(N+1)$ given
by:\footnote{This operator is in fact the superconformal primary of a short $B_1\overline{B}_1[0;0]^{(N+1,0)}$-multiplet \cite{Cordova:2016emh} (also sometimes referred as $\widehat{\mathcal{B}}_{N+1}$ in the nomenclature of \cite{Dolan:2002zh}).}%
\begin{equation}
\mathcal{O}_{\text{pure}}= \sqrt{\mathcal{Z}_N} X_{0}...X_{N},
\end{equation}
in the obvious notation. \ From this starting point, we can consider inserting
\textquotedblleft impurities,\textquotedblright\ by swapping out $X$'s for
$Y^{\dag}$'s. For example, we can insert one such impurity, leading to the
operator:%
\begin{equation}\label{Oi_definition}
	\mathcal{O}_{i}=\sqrt{\mathcal{Z}_N} \,X_{0}...X_{i-1}Y_{i}^{\dag}X_{i+1}...X_{N}\,.
\end{equation}
One can also consider adding further impurities, and provided the total
number is much smaller than $N$, we remain in the dilute gas approximation and
can treat the structure of correlation functions in a similar way. We first
focus on the case of a single impurity insertion since the
generalization to multiple impurities (at least in the dilute gas
approximation)\ follows a similar line of analysis.

We now proceed to the evaluation of operator mixing in this setup. In our
conventions, the two-point function for a free scalar will be normalized so
that:%
\begin{equation}
\left \langle (X_{i}^{\dag})_{B_{i}}^{A_{i+1}}(x)(
X_{i})_{B_{i+1}}^{A_{i}}(0) \right \rangle = \frac{\delta_{B_{i+1}%
}^{A_{i+1}}\delta_{B_{i}}^{A_{i}}}{\left\vert x\right\vert ^{2\Delta_{X}}},
\end{equation}
where $\Delta_X = 1$ for a 4D free field.
Here, the $A$ superscript and $B$ subscript indicate the
components of the bifundamental representation. With this convention, applying Wick's theorem to the two-point functions of $\mathcal{O}_i$ one ends up with traces over the indices of the gauge groups, each contributing a factor of $K$, so in this case the normalization factor is
$\mathcal{Z}_{N} = K^{-N}$.

To begin, we observe that if we switch off the gauge
couplings, the scaling dimension of each $\mathcal{O}_{i}$ is simply
$\Delta_{i}=(N+1)$. Once we switch on gauge interactions, we can expect the
two-point functions for the $\mathcal{O}_{i}$ to non-trivially mix. Our aim
will be to determine corrections to the two-point function:%
\begin{equation}
\left\langle \mathcal{O}_{i}^{\dag}(x)O_{j}(0)\right\rangle =\frac
{1}{\left\vert x\right\vert ^{2\Delta_{i}}}\times\left(  \delta_{ij}%
- \gamma_{ij}\log\left(  \left\vert x\right\vert^2 \Lambda^2 \right)%
+...\right)  .
\end{equation}
Here, $\gamma_{ij}$ refers to the matrix of anomalous dimensions. The main
idea is that in a basis of eigenoperators for the dilatation operator, we can
expect a shift in the scaling dimension, and this generates a logarithmic
correction term:%
\begin{equation}
\frac{1}{\left\vert x\right\vert ^{2(\Delta+\gamma)}}=\frac{1}{\left\vert
x\right\vert ^{2\Delta}}\times\left(  1 - \gamma\log\left(  \left\vert
x\right\vert^2 \Lambda^2 \right) +...\right)  .
\end{equation}

Let us now turn to operator mixing in this quiver gauge theory. We begin by
working to leading order in perturbation theory in the gauge couplings
$g_{i}^{2}$ but then show that the large R-charge limit allows us to form an
improved perturbation series. At first order in perturbation theory, the
operator mixing is controlled by the scalar potential of the $\mathcal{N}=2$
theory. To work this out, it is convenient to work in terms of the language of
4D $\mathcal{N}=1$ supersymmetry. In this case, each $X_{i}$ denotes a
bifundamental chiral multiplet and each $Y_{i}$ denotes a bifundamental in the
conjugate representation. Additionally, there is an adjoint-valued chiral
multiplet $Z_{i}$ for each gauge group. The scalar potential is a sum of
F-term and D-term contributions:%
\begin{equation}
V=V_{F}+V_{D}.
\end{equation}
For the D-term potential, this is just a sum over each of the gauge group
nodes:
\begin{gather}
V_{D}= \frac{1}{(4 \pi^2)^2}\underset{i=1}{\overset{N}{\sum}}\underset{a\in\text{adj}(G_{i})}{\sum
}\frac{1}{2}D_{i,a}^{2}\,,\\
D_{i,a}=g_{i}\left(  \text{Tr}_{i+1}(X_{i}^{\dag}T_{i,a}X_{i}-Y_{i}%
T_{i,a}Y_{i}^{\dag})-\text{Tr}_{i-1}(X_{i-1}T_{i,a}X_{i-1}^{\dag}%
-Y_{i-1}^{\dag}T_{i,a}Y_{i-1})\right)  + \cdots \,,
\end{gather}
where we have dropped contributions from the $Z_{i}$ scalars since they do not
contribute at leading order to operator mixing. The overall normalization by $1 / (4 \pi^2)^{2}$
is due to our normalization of the $X^{\dag} X$ two-point function.\footnote{With
canonically normalized kinetic terms for free fields, we would instead have a
two point function \begin{equation}
\left \langle (X_{i}^{\dag})_{B_{i}}^{A_{i+1}}(x)(
X_{i})_{B_{i+1}}^{A_{i}}(0) \right \rangle = \frac{1}{4 \pi^2} \frac{\delta_{B_{i+1}%
}^{A_{i+1}}\delta_{B_{i}}^{A_{i}}}{\left\vert x\right\vert ^{2\Delta_{X}}},
\end{equation} with $\Delta_X = 1$ for a 4D free field.}
Here, the $T_{i,a}$ are Lie
algebra generators for $G_{i}$, Tr$_{i+1}$ and Tr$_{i-1}$ indicate a
further instruction to sum over all of the indices of adjacent nodes in the
quiver. We also have the F-term potential contributions from the chiral
multiplet $Z_{i}$, which we denote as $F_{i,a}$:%
\begin{equation}
    V_{F}= \frac{1}{(4 \pi^2)^2} \underset{i=1}{\overset{N}{\sum}}\underset{a\in\text{adj}(G_{i})}{\sum
    }\left\vert F_{i,a}\right\vert ^{2}+ \cdots \,,
\end{equation}
where the \textquotedblleft$...$\textquotedblright\ refers to contributions
from the other chiral multiplets, which again do not influence operator mixing
at leading order. Here, we have:%
\begin{equation}
F_{i,a}=\sqrt{2}g_{i}\left(  \text{Tr}_{i+1}(Y_{i}T_{i,a}X_{i})-\text{Tr}%
_{i-1}(X_{i-1}T_{i,a}Y_{i-1})\right)  .
\end{equation}

From this, we can deduce that operator mixing is indeed possible. Since our
scalar potential only contains interaction terms between hypermultiplets on
neighboring links, we see that to leading order in perturbation theory the
operator $\mathcal{O}_{i}$, can only have a non-trivial two-point function
with $\mathcal{O}_{i-1}$, $\mathcal{O}_{i+1}$ and $\mathcal{O}_{i}$.

\begin{figure}
  \centering
  \includegraphics[scale=0.45]{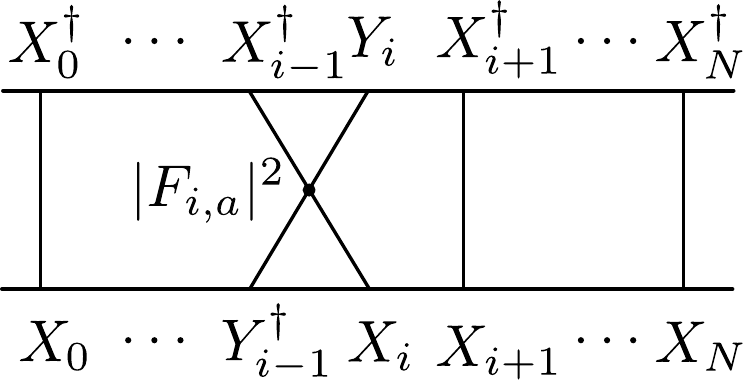}
  \caption{Diagram contributing at leading order to the correlation
  function of $\mathcal{O}^\dagger_i$ (top) and $\mathcal{O}_{i-1}$ (bottom) via an F-term interaction.}
\label{fig:diagram}
\end{figure}

The computation of the \textquotedblleft hopping terms\textquotedblright\
involves only a contribution from the F-term potential, see figure
\ref{fig:diagram}. In the case of hopping between $\mathcal{O}_{i}$ and
$\mathcal{O}_{i-1}$, the relevant contribution comes from $\left\vert
F_{i,a}\right\vert ^{2}$. We find the leading order contribution (evaluated in Euclidean
signature):
\begin{equation}\label{OOpopo}%
	\left<\mathcal{O}_{i}^{\dag}(x)\mathcal{O}_{i-1}(0)\right>
	=\frac{1}{\left| x\right|^{2\Delta_{0}}}\times\left(1 +  \frac{2g_{i}^{2} \widetilde{C}_{i}}{(4 \pi^2)^2}
	\int d^{4}z\frac{\left| x\right|^{4\Delta_X}}{\left| x-z\right|^{4\Delta_X}}
\frac{1}{\left| z\right| ^{4\Delta_X}} + \dots\right)\,,
\end{equation}
where we have suppressed flavor index structure.  Here, $\Delta_{0}$ refers to
the scaling dimension with all gauge couplings switched off and
$\widetilde{C}_{i+1}$ is a combinatorial factor obtained from the quadratic
Casimir of the fundamental representation and its dimension, given by
evaluating:%
\begin{equation}\label{eqn:blah}
  \widetilde{C}_{i}=\frac{1}{d_F}\underset{a\in\text{adj}(G_{i})}{\sum}\text{Tr}_{\text{fund}}%
(T_{i,a}T_{i,a})= \frac{f_{G_{i}}}{h_{G_{i}}^{\vee} d_F} \times \dim G_{i} \,,
\end{equation}
where $d_F$ is the dimension of the fundamental representation of $G_i$, and
with $h_{G}^{\vee}$ the dual Coxeter number of the group $G$.
The righthand side of (\ref{eqn:blah}) is obtained from evaluating traces in different
representations. To fix our normalizations we use the same notation as in
\cite{Ohmori:2014kda}, and introduce an auxiliary field strength $F$ to relate the various traces in different representations
as follows:
\begin{equation}
	\mathrm{Tr}_{\mathrm{fund}} F^2 = f_{G} \mathrm{Tr} F^2\,, \qquad
	\mathrm{Tr}_{\mathrm{adj}} F^2 = h_{G}^{\vee} \mathrm{Tr} F^2 \,.
\end{equation}
For a quiver with all $SU(K)$
factors, we have $f_{G}=1/2$ and $h_{G}^{\vee}=K$. In Appendix
\ref{app:LOOP} we evaluate
the loop integral appearing in (\ref{OOpopo}). Setting $\Delta_X = 1$, there is indeed
a logarithmic contribution, and we wind up with:%
\begin{equation}
\left\langle \mathcal{O}_{i}^{\dag}(x)\mathcal{O}_{i-1}(0)\right\rangle
=\frac{1}{\left\vert x\right\vert ^{2\Delta_{0}}}\left(1 +  \frac{g_{i}^2 \widetilde{C}_{i} }{8 \pi^2}
\times \log(\left\vert x\right\vert^2 \Lambda^2
) + ... \right) .
\end{equation}
Closely related to this case is the two-point function $\left\langle
\mathcal{O}_{i}^{\dag}(x)\mathcal{O}_{i+1}(0)\right\rangle $. The only
contribution comes from $\left\vert F_{i+1,a}\right\vert ^{2}$. The
computation is otherwise the same, so we get:%
\begin{equation}
\left\langle \mathcal{O}_{i}^{\dag}(x)\mathcal{O}_{i+1}(0)\right\rangle
=\frac{1}{\left\vert x\right\vert ^{2\Delta_{0}}}\times\left( 1 + \frac{g_{i+1}^2 \widetilde{C}_{i+1}}{8 \pi^2}
\times  \log(\left\vert x\right\vert^2
\Lambda^2) + ... \right) .
\end{equation}

Finally, we have the \textquotedblleft stationary term\textquotedblright%
\ associated with the evaluation of $\left\langle \mathcal{O}_{i}^{\dag
}(x)\mathcal{O}_{i}(0)\right\rangle $. This case is clearly somewhat more
subtle because there are far more contributions, which include contributions
both from the scalar potential, as well as t-channel vector boson exchange.
On general grounds, we expect there to be a zero mode in the anomalous dimension matrix
since we know that one linear combination of the $\mathcal{O}_i$ is in the same $SU(2)_{\mathcal{R}}$ R-symmetry
multiplet as $\mathcal{O}_{\mathrm{pure}}$. Jumping ahead to equation (\ref{eqn:gamma}),
we will indeed find that this is the case, provided we fix the relative strengths of the matrix entries.

In stringy terms we can argue as follows: starting from $K$ D3-branes probing $\mathbb{C}%
\times\mathbb{C}^{2}$, we can consider the related correlation functions
obtained from working in terms of the D3-brane probe of $\mathbb{C}%
\times\mathbb{C}^{2}/%
\mathbb{Z}
_{N+1}$. Doing so, we generate a circular quiver with $N+1$ gauge group nodes
$G_{i}=SU(K)$. We can arrive at this sort of quiver by gauging a diagonal
subgroup of $G_{0}\times G_{N+1}$. The evaluation of BMN-like operators is
thus quite close to our computation, and for the same reasons as presented
there, we conclude that the relevant contribution to the hopping term is
essentially the same, at least when all gauge couplings are equal. In that
setting, the contribution to the diagonal is a factor of $-2$ relative to the
off-diagonal hopping terms.

Another way to argue for this is to observe
that if we include all contributions other than those from the F-term
contributions, we expect an exact cancellation, much as in the computations of
reference \cite{Gadde:2010zi}. That leaves us to include the
contributions from $\left\vert F_{i,a}\right\vert ^{2}$ and $\left\vert
F_{i+1,a}\right\vert ^{2}$. This leads us to our formula:
\begin{equation}\label{eqn:oneloop}
\left\langle \mathcal{O}_{i}^{\dag}(x)\mathcal{O}_{i}(0)\right\rangle
=\frac{1}{\left\vert x\right\vert ^{2\Delta_{0}}}\times\left(  1 - \frac{1}{8 \pi^2}\left(
  g_{i}^{2}\widetilde{C}_{i} + g_{i+1}^{2}\widetilde{C}_{i+1}\right)  \log(\left\vert
x\right\vert^2 \Lambda^2) + ... \right)  .
\end{equation}

Summarizing, we have extracted the anomalous dimension matrix with non-zero
entries:%
\begin{align}
  \gamma_{ii} &
  = \frac{1}{8 \pi^2} \left(g_{i}^{2}\widetilde{C}_{i}+ g_{i+1}^{2}\widetilde{C}_{i+1} \right)\\
  \gamma_{i,i+1} &  =\gamma_{i+1,i}=- \frac{g_{i+1}^{2}\widetilde{C}_{i+1}}{8 \pi^2},
\end{align}
for $i=0,...,N$. Setting $g_{i}^{2}=g^{2}$ for $i=1,...,N$ and $g_{0}%
^{2}=g_{N+1}^{2}=0$, we can also write this as a 1D Lattice Laplacian with
non-trivial boundary conditions:%
\begin{equation} \label{eqn:gamma}
\underline{\gamma}= \lambda_{G}  \left[
\begin{array}
[c]{ccccc}%
1 & -1 &  &  & \\
-1 & 2 & -1 &  & \\
& -1 & ... & -1 & \\
&  & -1 & 2 & -1\\
&  &  & -1 & 1
\end{array}
\right]  ,
\end{equation}
where $\lambda_{G}$ is the overall normalization for the Hamiltonian and is given by:
\begin{equation}
\lambda_{G} = \frac{g^2 \widetilde{C}_{G}}{8 \pi^2},
\end{equation}
for a general gauge group $G$, with $\widetilde{C}_G$ as in equation
(\ref{eqn:blah}) evaluated in the special case $G = SU(K)$. Note that
equation (\ref{eqn:gamma}) is in accord with having a zero mode.

It is also instructive to extract the linear combination of $\mathcal{O}_{i}%
$ which are diagonal under the action of the anomalous dimension matrix.
We find operators $\widetilde{\mathcal{O}}_{p}$ with scaling dimensions:
\begin{align} \label{magnon}
	\widetilde{\mathcal{O}}_{p} &  = \frac{1}{\sqrt{N+1}}\underset{j=0}{\overset{N}{\sum}}\cos\left(p\left(j+\frac{1}{2}\right)  \right)  \mathcal{O}_{j}\\
	(\Delta-\Delta_{0})_{p} & = 4 \lambda_{G} \sin^{2}\left(  \frac{p}{2}\right)  \text{ \ \ with \ }\lambda_{G}= \frac{g^{2}}{8 \pi^2} \frac{f_{G}\dim G}{h_{G}^{\vee} d_{F}}
\end{align}
where the original dimension is $\Delta_{0}=(N+1)$ and the eigenoperator,
$\widetilde{O}_p$, is interpreted as a magnon with quantized momentum, $p=\frac{\pi
m}{N+1}\,,m=0,1,\dots,N$. These $\widetilde{O}_p$ are the eigenoperators for the operator mixing captured by the one-loop two-point function given in (\ref{eqn:oneloop}). The important point for us is that just as in
\cite{Berenstein:2002jq}, even though we performed a perturbative expansion in
$g^{2}$, we can expand at large $N$. In this limit, we observe that even if
$g^{2}$ becomes large, we can continue to work to leading order in
perturbation theory provided $g^{2}/N^{2}$ remains small.

\subsection{A Spin Chain}

In the above analysis we focused on the special case of a single impurity insertion and its motion throughout the
spin chain. We can generalize this to additional impurities as associated with the more general class of operators:
\begin{equation}
\mathcal{O}_{m_0,...,m_N} = \sqrt{\mathcal{Z}_N} X_{0}^{(m_0)} ... X_{N}^{(m_N)}
\end{equation}
where $m_{i} = \pm 1/2$ since we are dealing with the spin $1/2$ representation of $SU(2)_{\mathcal{R}}$ R-symmetry on each spin site.
The main idea is to associate this operator with a corresponding configuration of spins in a 1D spin chain:
\begin{equation}
\vert m_0 , ..., m_N \rangle \leftrightarrow \mathcal{O}_{m_0,...,m_N}.
\end{equation}
Operator mixing is controlled by the spin chain Hamiltonian with open boundary conditions:
\begin{equation}
  H_{A} = - \lambda_{A} \sum_{i=0}^{N-1} \left(2 \overrightarrow{S}_{i} \cdot \overrightarrow{S}_{i+1} - \frac{1}{2} \right),
\end{equation}
where we have introduced the spin $s = 1/2$ operators $\overrightarrow{S}_{i}$
for each spin site. Here, the coefficient $\lambda_{A}$ is the same as in line (\ref{magnon}).
Indeed, the specific numerical pre-factor has been chosen so that we recover the dispersion relation
$E_{A} = \lambda_{A} p^2$ in the small momentum limit. The ground state energy has been fixed so that it is zero.
This is in accord with the fact that $1/2$-BPS operators
do not receive a correction to their anomalous dimension.

We observe that this is just the Hamiltonian of the celebrated
Heisenberg $XXX_{s = 1/2}$ spin chain, with open boundary conditions. This is a well-known
integrable system and the spectrum of excitations can be studied in the standard way. As a preliminary comment,
we simply observe that since the operator:
\begin{equation}
\overrightarrow{S}_{\mathrm{tot}} = \underset{i}{\sum} \overrightarrow{S}_{i}
\end{equation}
commutes with $H_{A}$, we can perform a block diagonalization of operator mixing so that the total number of impurities remains constant.
Our analysis of single impurity insertions clearly generalizes.

To further analyze the resulting spectrum excitations, we
now turn to the Bethe Ansatz for the open ferromagnetic Heisenberg spin chain.
In the sector with $I$ impurities, we have corresponding quasi-particle momenta $p_{1},...,p_{I}$.
These are conveniently expressed in terms of complex rapidities $\mu
_{1},...,\mu_{I}$ which are related to the quasi-particle momenta as:%
\begin{equation}\label{eqn:rapidity}
\exp(ip_{j})=\frac{\mu_{j}+\frac{i}{2}}{\mu_{j}-\frac{i}{2}}.
\end{equation}
The specific values of the rapidities are fixed by the Bethe ansatz equations
of the open spin chain \cite{WANG}:
\begin{equation}
\left(  \frac{\mu_{j}+\frac{i}{2}}{\mu_{j}-\frac{i}{2}}\right)  ^{2(N+1)}%
=  \underset{l\neq j}{%
{\displaystyle\prod}
}\frac{\left(  \mu_{j}-\mu_{l}+i\right)  \left(  \mu_{j}+\mu_{l}+i\right)
}{\left(  \mu_{j}-\mu_{l}-i\right)  \left(  \mu_{j}+\mu_{l}-i\right)  },
\end{equation}
for $j=1,...,I$ impurity excitations.

The energy (i.e. anomalous dimension) of a given eigenoperator is then given by:
\begin{equation}
(\Delta - \Delta_{0}) = E_{A}=\lambda_{A}\underset{j=1}{\overset{I}{%
{\displaystyle\sum}
}} \epsilon(\mu_j),
\end{equation}
where $\Delta_{0} = N+1$ and the energy of a given quasi-particle excitation is:
\begin{equation}
\epsilon(\mu) = \frac{i}{\mu + \frac{i}{2}} - \frac{i}{\mu - \frac{i}{2}}.
\end{equation}
It is appropriate to refer to the $\mu_i$ as rapidities because we have a dispersion relation of the form:
\begin{equation}
\epsilon(\mu) = \frac{d}{d \mu} p(\mu),
\end{equation}
where $\epsilon(\mu)$ is proportional to the energy of a quasi-particle excitation.

In fact, all normalizations are fixed once we specify the behavior of the single impurity excitations. To see this, consider expanding
equation (\ref{eqn:rapidity}) at small $p$ / large $\mu$. This leads to the relation:
\begin{equation}
p \simeq \frac{1}{\mu} + ...
\end{equation}
Plugging in our expressions, we get:
\begin{equation}
\epsilon(p) \simeq p^2 + ....
\end{equation}

Note that the Bethe ansatz equations tell us about the spectrum of excitations \textit{above} the ground state. The ground state itself 
is associated with BPS operators of the 6D SCFT, and as such have precisely zero momentum in the spin chain. We emphasize that the $\mu \rightarrow \infty$ limit must be treated separately from the rest of the spectrum of excitations, where $\mu$ is finite.

As a last comment, we expect that at higher order in perturbation theory that
each loop order allows a spin to interact with neighbors further away.
Even so, we expect the structure of integrability to persist in some form.

\subsection{More General Spin Chains}

Though we have focused on adding impurities to the ``ground state'', $\mathcal{O}_{\mathrm{pure}} = \sqrt{\mathcal{Z}_{N}} X_0 ... X_N$,
it is clear that we could also consider a broader class of operators. As an example of this sort,
consider the class of operators obtained from inserting $(Y_{i}X_{i})$ such as:%
\begin{equation}
	\mathcal{B}_{i}=\sqrt{\mathcal{Z}_{\mathcal{B}_i}} X_{0}...X_{i-1}X_{i}(Y_{i}X_{i})X_{i+1}...X_{N}.
\end{equation}
This is superficially quite similar to the operators $\mathcal{O}_{i}$ just
considered, and we can again see that F-term exchange leads to a hopping term
for the location of the impurity. Note that in this case, we again have some
protection against various operator mixing effects since, for example $X$ and
$Y$ have the same charge under the Cartan subalgebra of $\mathfrak{su}(2)_{R}$.
Observe, however, that this operator can also mix with another class of operators in which a
mesonic operator has ``bubbled off''. For example, there appears to be nothing which prevents mixing
with the operator:
\begin{equation}
\mathcal{B}^{\text{bubble}}_{i}=\sqrt{\mathcal{Z}_{\mathcal{B}^{\text{bubble}}_i}} X_{0}...X_{i-1}X_{i}X_{i+1}...X_{N} \times \mathrm{Tr}(Y_{i} X_{i}).
\end{equation}
This sort of mixing can be suppressed provided we work in the limit where the ranks of gauge groups are also large. Standard results in large $K$ gauge theory, see for instance the review \cite{Aharony:1999ti}, demonstrate that such ``multi-trace'' operators are suppressed by additional powers of $K$.

As another example, consider the set of operators which form a closed
loop beginning at a gauge group site $i$ and extending out $L$ gauge group
sites:%
\begin{equation}
	\mathcal{C}_{i,i+L}= \sqrt{\mathcal{Z}_{\mathcal{C}_{i,i+L}}}\text{Tr}(X_{i}...X_{i+L}Y_{i+L}...Y_{i}).
\end{equation}
In this case, we have a flavor neutral operator and we can see the same sort of
``bubbling off'' of mesons, which leads to non-trivial operator mixing with multi-trace operators.
Note that because in this case we have only closed loops, there is less protection from operator mixing, and so there can be a transition to an operator such as:
\begin{equation}
\mathcal{C}^{\text{bubble}}_{i,i+L}= \sqrt{\mathcal{Z}_{\mathcal{C}^{\text{bubble}}_{i,i+L}}}\text{Tr}(X_{i}Y_{i})...\mathrm{Tr}(X_{i+L}Y_{i+L}).
\end{equation}
Again, we can suppress this in the limit where the ranks of the gauge groups are large, in which
case we expect the ``ground state'' to be approximately protected from such bubbling. In this limit
we can also consider adding impurity insertions, and from this we can extract a quite
similar analysis of hopping terms.

\subsection{More General Quivers}

We can also contemplate a more general class of quivers with different gauge
group ranks. The condition that we retain a conformal field theory is that the
beta function for each gauge coupling vanishes. One
possibility is a circular quiver with gauge group
$SU(K)^{N+1}$. We remark that this sort of quiver shows up with $K$ D3-branes
probing a $\mathbb{C}^{2}/%
\mathbb{Z}
_{N}$ singularity \cite{Douglas:1996sw}, and also arises from the 6D\ SCFT obtained from $K$
M5-branes probing a $\mathbb{C}^{2}/%
\mathbb{Z}
_{N}$ singularity, compactified on a further $T^{2}$ (see e.g. \cite{DelZotto:2014hpa}).
Note also that if we had instead moved onto the tensor branch of this 6D\ SCFT\ and then compactified,
we would have arrived at the T-dual theory with gauge group $SU(N)^{K+1}$
arranged in a linear quiver. The two theories are related by T-dualities /
flavor Wilson lines in the $T^{2}$ direction.

Another way to get a more general class of quiver gauge theories is achieved
by adjusting the individual ranks of our linear quiver. To maintain
conformality on each gauge group node we must introduce some additional
flavors $M_i$ in the fundamental representation. The condition on the gauge
group nodes is then:%
\begin{equation}
2K_{i}-K_{i-1}-K_{i+1}= M_{i}\geq0.
\end{equation}
This leads to a strictly convex profile for the $K_{i}$, with a maximum
plateau possible in the middle of the quiver. While a full analysis of the
spectrum of operator mixing in this class of theories is clearly somewhat more
challenging, we can already see that the matrix of anomalous dimensions again
resembles a 1D\ spin chain Hamiltonian, but now with more non-trivial boundary
conditions on the left and right. To see this, consider again operator mixing
for the $\mathcal{O}_{i}$ operators with a single impurity insertion.
Eigenoperators of the dilatation operator satisfy the eigenvalue equation:%
\begin{equation}
\gamma_{ij}v_{j}=\kappa v_{j}. \label{eigenequation}%
\end{equation}
Returning to our general expression for operator mixing in these theories, we
have:%
\begin{align}
  \gamma_{ii}  &
  = \frac{1}{8 \pi^2} \left( g_{i}^{2}\widetilde{C}_{i} + g_{i+1}^{2}\widetilde{C}_{i+1} \right) \\
  \gamma_{i,i+1}  &  =\gamma_{i+1,i}=-\frac{g_{i+1}^{2}\widetilde{C}_{i+1}}{8 \pi^2},
\end{align}
for $i=0,...,N$. Setting $g_{i}^{2}=g^{2}$ for $i=1,...,N$ and $g_{0}%
^{2}=g_{N+1}^{2}=0$, we can also write the eigenvalue equation as a 1D Lattice
Laplacian with non-trivial boundary conditions. For example, in the
\textquotedblleft middle region\textquotedblright\ where all the gauge group
ranks are the same, we just have the condition:%
\begin{equation}
	\frac{g^{2}}{8 \pi^2} \widetilde{C}\times(2v_{j}-v_{j+1}-v_{j-1})=\kappa v_{j},
\end{equation}
which, in the continuum limit available by taking large operator scaling
dimensions, takes the form:%
\begin{equation}
  -\frac{g^{2}}{8 \pi^2} \widetilde{C}\times\frac{d^{2}v(x_{\bot})}{d
  x_{\bot}^{2}}=\kappa v(x_{\bot}),
\end{equation}
for $x_{\bot}$ a coordinate along the line of gauge groups on the tensor branch.
Now, as we approach the regions with varying gauge groups, we observe a more
general eigenvalue equation which we can write as:%
\begin{equation}
  \frac{g^{2}}{8 \pi^2} \left(  -\widetilde{C}_{1}\left(  x_\bot \right)  \times\frac{d^{2}v(x_\bot)}{d x_{\bot}^{2}%
    }-\widetilde{C}_{2}\left(  x_\bot \right)  \times\frac{dv( x_\bot)}{d
    x_\bot}-\widetilde{C}_{3}\left(  x_\bot \right)
v(x_\bot)\right)  =\kappa v(x_\bot),
\end{equation}
for suitable convex position dependent profiles for the $\widetilde{C}_{1,2,3}(x_{\bot})$. It would be
very interesting to study the resulting profile of operator mixing effects as
a function of the different choices, but we defer this to future work.

\section{6D SCFTs with Classical Matter \label{sec:6DATYPE}}

In the previous section we focused on the appearance of a spin chain sector of
some 4D $\mathcal{N} = 2$ SCFTs. In this section we turn to a quite similar analysis for
6D\ SCFTs with classical matter. More precisely, we consider a class of
6D\ SCFTs with tensor branch given by a classical quiver gauge theory. The
specific case of interest in this section will be quivers of the form:%
\begin{equation}
\lbrack G_{0}]-G_{1}-...-G_{N}-[G_{N+1}],
\end{equation}
where here, each gauge group factor corresponds to $G_{i}=SU(K)$. In this
case, we have bifundamental hypermultiplets $X_{i}\oplus Y_{i}^{\dag}$ between
each link. Each gauge group factor also pairs with a tensor multiplet with
scalar vev controlling the value of the gauge coupling. The vev of the tensor
multiplet scalar has dimensions of mass$^{2}$ and controls the tension of a
$1/2$-BPS\ string in the 6D\ effective field theory.

We can deduce the precise relation between the gauge coupling and this tension
by noting that such strings also arise as solitonic excitations in the
6D\ gauge theory. Using any number of string theory realizations, we can then
extract the relation between the gauge coupling of the 6D field theory and
this string tension which we can also set equal to the vev of a tensor multiplet
scalar (in our normalizations):\footnote{For example, in a setup where we
engineer the 6D\ SCFT\ using D5-branes probing an A-type singularity
\cite{Blum:1997mm}, the gauge coupling of the D5-brane is set by the tension
of the D5-brane by expanding the DBI\ action (see e.g. \cite{Polchinski:1998rr}):%
\begin{equation}
\frac{1}{4g^{2}}=\frac{1}{2\pi g_{s}}\frac{1}{\left(  2\pi\ell_{s}\right)
^{2}}.
\end{equation}
where $g_{s}$ denotes the string coupling and $\ell_{s}$ is the string length.
The solitonic excitation is associated with a D1-brane filling a
2D\ spacetime. This comes with a tension of:%
\begin{equation}
T_{D1}=\frac{2\pi}{g_{s}}\frac{1}{\left(  2\pi\ell_{s}\right)  ^{2}}%
=\frac{(2\pi)^{2}}{4g^{2}}.
\end{equation}
So, in what follows we introduce tensor multiplet scalars $T_{i}$ normalized
so that:%
\begin{equation}
\left\langle T_{i}\right\rangle =\frac{(2\pi)^{2}}{4g_{i}^{2}}.
\end{equation}
}%
\begin{equation}
\left\langle T_{i}\right\rangle =\frac{(2\pi)^{2}}{4g_{i}^{2}}.
\end{equation}

Now, in this 6D\ effective field theory, we can construct a similar class of
operators to those studied in section \ref{sec:4DATYPE}. Of course, since this
is not a conformal field theory, we should not expect that our analysis of
hopping terms will carry through. One symptom of this is that our gauge
coupling is now dimensionful, and to reach a fixed point we will need to
extrapolate this to strong coupling at the origin of the tensor branch.

But from what we have seen in the previous section, we can anticipate that a
perturbative expansion may nevertheless be available for some subsectors of
operators. Indeed, provided the scaling dimension of a candidate
operator is large as well, we can hope that a perturbative expansion will
still be available. This is essentially the argument of \cite{Berenstein:2002jq},
but now applied to 6D\ SCFTs.

We now argue that a perturbative expansion is still available. To see this,
consider the holographic dual of our SCFT, as obtained from $N$ M5-branes
probing a $\mathbb{C}^{2}/%
\mathbb{Z}
_{K}$ singularity. In this limit, we reach the geometry $AdS_{7}\times S^{4}/%
\mathbb{Z}
_{K}$ with $N$ units of four-form flux threading the $S^{4}/%
\mathbb{Z}
_{K}$. This geometry comes with two orbifold fixed points at the north and
south poles of the sphere, and so the gravity dual is actually coupled to a
pair of 7D\ super Yang--Mills theories with gauge group $SU(K)$. Now, an
interesting feature of this geometry is that the tensor branch description
literally \textquotedblleft deconstructs\textquotedblright\ a great arc which
passes from the north pole to the south pole. With this in mind, we can
consider the effect of being slightly on the tensor branch as actually
registering some fine-grained structure in the holographic dual. To
corroborate this picture, consider moving slightly onto the tensor branch. In
the holographic dual this means we separate the M5-branes both down the
throat of the $AdS_{7}$ geometry, and also means they are separated at
different longitudes of $S^{4}/%
\mathbb{Z}
_{K}$. Working in the limit where each M5-brane is uniformly separated from
its neighbors, the arc length for a sphere of radius $R_{S^{4}/%
\mathbb{Z}
_{K}}=\ell_{p}\left(  \pi N\right)  ^{1/3}$ (see \cite{Maldacena:1997re})
gives us $N$ equal segment pieces, each of length:%
\begin{equation}
\frac{\pi R_{S^{4}/%
\mathbb{Z}
_{K}}}{N}=\frac{\pi^{4/3}\ell_{p}}{N^{2/3}}.
\end{equation}
Wrapping an M2-brane over one such segment leads to a 6D\ effective string in
the tensor branch theory. The tension of this effective string is:%
\begin{equation}
T_{\text{eff}} = \frac{\pi^{4/3}\ell_{p}}{N^{2/3}}\times\frac{2\pi}{\left(
2\pi\ell_{p}\right)  ^{3}}=\frac{(2 \pi)^2}{4g_{6D}^{2}}, \label{TeffMtheory}%
\end{equation}
where we have also written the 6D\ gauge coupling as obtained from
compactifying our 7D\ super Yang--Mills theory on the interval. Now we can see
that, at least for states with sufficiently large mass, a perturbative
expansion may be available.

\begin{figure}
  \centering
  \includegraphics[scale=0.35]{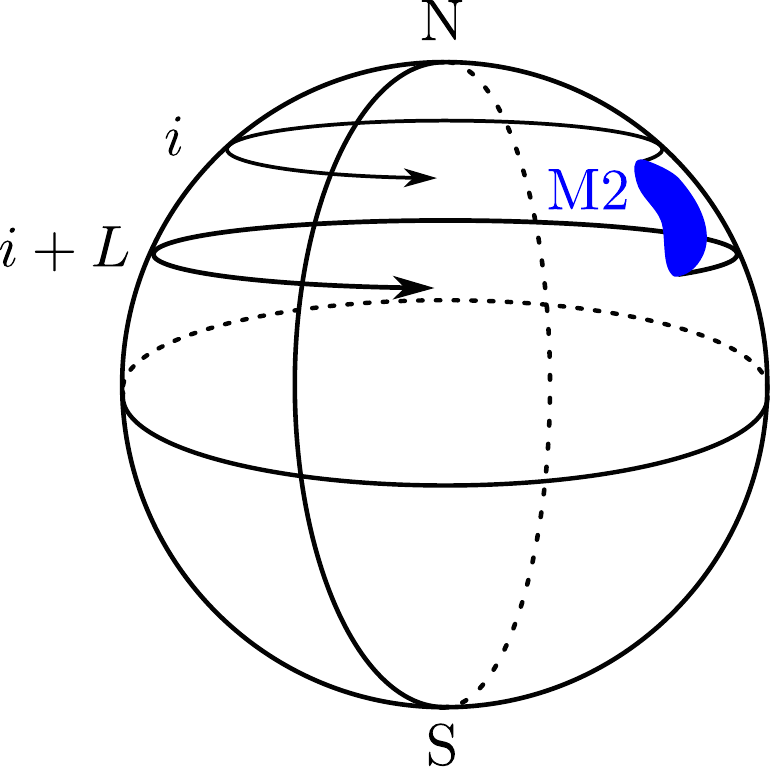}
  \caption{Orbiting M2-brane wrapped on an $S^2\subset (S^4/\mathbb{Z}_K)$ stretched between
  constant latitudes $i$ and $i+L$. The associated states give rise to the operators $\mathcal{C}_{i,i+L}$ in the CFT.}
\label{fig:M2S4}
\end{figure}

Indeed, starting from the geometry $AdS_{7}\times S^{4}/%
\mathbb{Z}
_{K}$ we can take a pp-wave limit, much as in
reference \cite{Berenstein:2002jq} (see also \cite{Berenstein:2002zw}).
Then, performing discrete light cone quantization along a circle of radius
$R_{S^{4}/%
\mathbb{Z}
_{K}}=\ell_{p}\left(  \pi N\right)  ^{1/3}$, we get operators in the dual CFT of
scaling dimension $\Delta\sim N^{1/3+\varepsilon}$. Fluctuations in the
spectrum of graviton excitations translate in the dual CFT to perturbations in
the scaling dimension of operators such as:%
\begin{equation}
\mathcal{C}_{i,i+L}= \sqrt{\mathcal{Z}_{\mathcal{C}_{i,i+L}}} \text{Tr}(X_{i}...X_{i+L}Y_{i+L}...Y_{i}),
\end{equation}
as well as fluctuations generated by impurity insertions. This sort of
operator corresponds to an M2-brane wrapped on an $S^{2}$ which is orbiting
along a circle trapped on a specific latitude of $S^{4}/%
\mathbb{Z}
_{K}$, see figure \ref{fig:M2S4}. Note that
the identification with a single trace operator
is only approximately true due to ``mesonic bubbling,'' but this
becomes more accurate if we assume a suitable large $K$ / planar approximation
(see e.g. \cite{Witten:1998xy, Balasubramanian:2001nh}).
The brane does not collapse because it has non-zero angular momentum.
Here, the precise value of $i$ indicates the northern latitude on the $S^{4}/%
\mathbb{Z}
_{K}$, and $L$ controls the overall angular momentum / size of the object. To
see controlled perturbations in the holographic dual, we would need to take
$L\sim N^{1/3}$, but we can (and will) consider faster scaling in $L$. In
those cases, the spectrum of perturbations will be washed out to leading order
in the holographic dual.

Having seen that a perturbative expansion should indeed be available, we now
turn to a direct analysis focused on the structure of the 6D\ theory itself.
We have already noted that a dimensionless perturbation parameter is available
for operators with large R-charge, so provided we can suitably regulate our
6D\ theory, we should expect to be able to carry out computations. Our main
proposal for doing this is to try and recast the gauge coupling in terms of a
dimensionless parameter. As we have already mentioned, in the case of the 4D
computation, the relevant \textquotedblleft hopping terms\textquotedblright%
\ are controlled by the scalar potential. In 6D, something similar holds, and
we have a triplet of D-term constraints. On the partial tensor branch, 
$\mathcal{N}=(1,0)$ supersymmetry requires that the effective potential for the hypermultiplets is a sum of squares for the triplet of 
moment maps, namely $V \sim g_{6D}^2 (\vert D_1 \vert^2 + \vert D_2 \vert^2 + \vert D_3 \vert^2)$, which we write 
schematically as:
\begin{equation}
V_{6D}\sim g_{6D}^{2}\left\vert X_{6D}\right\vert ^{4}.
\end{equation}
Since we have free hypermultiplets, we can assign each a scaling dimension of
$\Delta_{6D}=2$, which is in accord with the fact that the scalar of the tensor multiplet 
has scaling dimension $\Delta_{6D}=2$. Now, observe that if we
compactify this theory on a circle each 6D hypermultiplet becomes a 5D
hypermultiplet, with the relation:%
\begin{equation}
\left(  2\pi\ell_{KK}\right)  ^{1/2}X_{6D}=X_{5D}.
\end{equation}
Plugging back into $V_{6D}$, we obtain:%
\begin{equation}
V_{6D}\sim\left(  2\pi\ell_{KK}\right)  ^{-2}g_{6D}^{2}\left\vert
X_{5D}\right\vert ^{4},
\end{equation}
so in terms of the combination $\left(  2\pi\ell_{KK}\right)  ^{-2}g_{6D}^{2}$
we indeed have a dimensionless parameter. To get a 6D\ answer, we should
really view $X_{5D}$ as a collection of 5D\ fields labelled by points along
the compactification circle. So, we are really performing a computation in the
5D\ KK\ theory in which we retain all of the Kaluza--Klein modes associated
with dimensional reduction. We will refer to this as a \textquotedblleft%
5D\ KK\ regulator\textquotedblright\ since it involves a computation in this theory.

Now, to actually extract a number for operator mixing from this process we
will also need to find a way to relate the scales associated with $\left(
2\pi\ell_{KK}\right)  ^{-2}$ and $g_{6D}^{2}$. To do so, we again appeal to
the M-theory / holographic dual description. In the directions transverse to
the M5-branes, we have identified a minimal length scale of separation, as set
implicitly by equation (\ref{TeffMtheory}). The non-trivial $N$-scaling can be
attributed to the backreaction of the M5-branes on the geometry. In the
directions along the M5-brane, however, we expect that reduction on a
Planckian circle of volume $2\pi\ell_{p}$ is the minimal length scale
available for reduction. Putting these relations together, we get an effective
dimensionless coupling:%
\begin{equation}
g_{\text{eff}}^{2}=\left(  2\pi\ell_{KK}\right)  ^{-2}g_{6D}^{2}= (N+1)^{2/3} \pi^{2/3}, \label{gefftwo}%
\end{equation}
which is dimensionless, but also quite large. In the above, we have written the formula
for $N+1$ M5-branes since this is the convention used in our discussion of generalized quivers.

With this in place, we can now proceed to an analysis of operator mixing for
certain subsectors. Our plan will be to essentially follow the same line of
analysis presented in our discussion of 4D\ quivers, with the proviso that
now, our loop integrals must be performed in the 5D\ KK\ regulated theory.

\subsection{A Spin Chain}

As a first example, consider the $1/2$-BPS\ operator given on the tensor
branch by:
\begin{equation}
\mathcal{O}_{\text{pure}}= \sqrt{\mathcal{Z}_N} X_{0}...X_{N},
\end{equation}
in the obvious notation. This operator is in the bifundamental representation
of $SU(K)_{L}\times SU(K)_{R}$, just as in the 4D\ case. In 6D, the scaling
dimension at the conformal fixed point is $\Delta_\text{pure}=2(N+1)$ since
each $\Delta_{X}=2$. In terms of the nomenclature introduced in
\cite{Cordova:2016emh, Buican:2016hpb}, which we briefly review in Appendix
\ref{app:superRepresentations}, this operator is the superconformal primary of
a type $\mathcal{D}[0,0,0]^{\frac{N+1}{2}}_{2(N+1)}$ multiplet.

We note that the existence of this operator at the
conformal fixed point as well as its scaling dimension is in accord with our
discussion of brane recombination given in section \ref{sec:GENQUIV}.

Starting on the tensor branch, we again consider inserting an impurity. For
example, we can insert one such impurity. This leads to operators such as:%
\begin{equation}
	\mathcal{O}_{i}=\sqrt{\mathcal{Z}_{N}} X_{0}...X_{i-1}Y_{i}^{\dag}X_{i+1}...X_{N}%
\end{equation}
One can also consider adding further impurities, and provided the total number
is much smaller than $N$, we remain in the dilute gas approximation and can
treat the structure of correlation functions in a similar way.

Let us now proceed to study operator mixing in this theory. The calculation is
essentially the same as that for the 4D theory; we have a triplet of D-terms
which contributes to the hopping term and to the \textquotedblleft stationary\textquotedblright%
\ term. For example, we can evaluate the hopping term in the
5D\ KK-regulated theory (we work in Euclidean signature):\footnote{In our
conventions, the normalization of the two-point function for a 6D free field is:
\begin{equation}
\left \langle (X_{i}^{\dag})_{B_{i}}^{A_{i+1}}(x)(
X_{i})_{B_{i+1}}^{A_{i}}(0) \right \rangle = \frac{\delta_{B_{i+1}%
}^{A_{i+1}}\delta_{B_{i}}^{A_{i}}}{\left\vert x\right\vert ^{2\Delta_{X}}},
\end{equation}
with $\Delta_X = 2$ for a free field. In the 5D
KK regulated theory we replace each propagator appearing
in the loop integral with $\Delta_X = 3/2$. Note also that with canonically normalized kinetic
terms for free fields,
the two-point function would be:
\begin{equation}
\left \langle (X_{i}^{\dag})_{B_{i}}^{A_{i+1}}(x)(
X_{i})_{B_{i+1}}^{A_{i}}(0) \right \rangle = \frac{1}{4 \pi^3} \frac{\delta_{B_{i+1}%
}^{A_{i+1}}\delta_{B_{i}}^{A_{i}}}{\left\vert x\right\vert ^{2\Delta_{X}}}.
\end{equation}
}
\begin{equation}
\left\langle \mathcal{O}_{i}^{\dag}(x)\mathcal{O}_{i-1}(0)\right\rangle
=\frac{1}{\left\vert x\right\vert ^{2\Delta_{0}}}\times\left(1 +  \frac{2g_{i}^{2}%
  \widetilde{C}_{i}}{(4 \pi^{3})^{2}} \int d^{6}z\frac{\left\vert x\right\vert ^{4\Delta_{KK}}}{\left\vert
x-z\right\vert ^{4\Delta_{KK}}}\frac{1}{\left\vert z\right\vert ^{4\Delta
_{KK}}}\right)  ,
\end{equation}
where here, $\Delta_{KK}=3/2$, the scaling dimension of a free hypermultiplet in a
5D\ SCFT. Additionally, each $g_{i}^{2}$ is specified as in equation
\eqref{gefftwo}. We evaluate this integral in Appendix B, obtaining:%
\begin{equation}
\left\langle \mathcal{O}_{i}^{\dag}(x)\mathcal{O}_{i-1}(0)\right\rangle
=\frac{1}{\left\vert x\right\vert ^{2\Delta_{0}}}\times\left(1 + \frac{g_{i}^{2}\widetilde{C}_{i}}{16 \pi^3} \times \log(\left\vert x\right\vert^2 \Lambda^2) + ...  \right).
\end{equation}
Let us make a few comments here. First, we observe that as expected, we
achieve a logarithmic correction to the two-point function, in accord with the
interpretation of a small shift in the anomalous dimension matrix.
Additionally, we note that this would not have worked if we had set
$\Delta_{KK}$ to the 6D\ scaling dimension of our fields. This provides an a
posteriori justification for our regulator. Lastly, we note that the strength
of the gauge coupling is quite large, so we must indeed work at large R-charge
to extract a perturbative contribution to the mixing matrix. Again, being at large R-charge is vital to avoid any issues with the one-loop perturbative computation as discussed at the beginning of section \ref{sec:4DATYPE}.

So, much as in the 4D case we get operator mixing on the tensor branch dictated by the
matrix $\gamma_{ij}$ with non-zero entries:
\begin{align}
  \gamma_{ii} &
  =\frac{1}{16 \pi^{3}} \left( g_{i}^{2}\widetilde{C}_{i} + g_{i+1}^{2}\widetilde{C}_{i+1} \right)\\
  \gamma_{i,i+1} &  =\gamma_{i+1,i}=- \frac{g_{i+1}^{2}\widetilde{C}_{i+1}}{16 \pi^3} \,,
\end{align}
for $i=0,...,N$, and all other entries vanish. Setting $g_{i}^{2}=g_{\mathrm{eff}}^{2}$ for $i=1,...,N$ and $g_{0}%
^{2}=g_{N+1}^{2}=0$, we can also write this as a 1D Lattice Laplacian with
open boundary conditions:%
\begin{equation}
\underline{\gamma}= \lambda_{A}
\left[
\begin{array}
[c]{ccccc}%
1 & -1 &  &  & \\
-1 & 2 & -1 &  & \\
& -1 & ... & -1 & \\
&  & -1 & 2 & -1\\
&  &  & -1 & 1
\end{array}
\right]  ,
\end{equation}
where
\begin{equation}
\lambda_{A} = \frac{g_{\mathrm{eff}}^2 \widetilde{C}_{A}}{16 \pi^3},
\end{equation}
with $g_{\mathrm{eff}}^2 = (N+1)^{2/3} \pi^{2/3}$, and we have the group theory factor:
\begin{equation}
  \widetilde{C}_{G}=\frac{f_{G}}{h_{G}^{\vee} d_F}\times\dim G \,,
\end{equation}
with $d_F$ the dimension of the fundamental representation of the A-type gauge group $G$.
Again, operator mixing is dictated by a spin chain Hamiltonian:
\begin{equation}
H_{A} = - \lambda_{A} \underset{i}{\sum} \left(2 \overrightarrow{S}_{i} \cdot \overrightarrow{S}_{i+1} - \frac{1}{2} \right).
\end{equation}
The main difference from the 4D case is that the constant $\lambda_{A}$ is now fixed by a one loop
computation in the 5D KK-regulated theory.

Now, in spite of these similarities with the 4D case, we also note that the operators
we have been studying are really specified on the tensor branch. Indeed, we now need to
take a decoupling limit so that the transverse momentum $P_{\bot} = 0$, as
per our discussion in subsection \ref{ssec:DECOUP}. At least in the single impurity sector, this removes
\textit{all but one} of the operators, and we are left with the single zero mode:
\begin{equation}
\widetilde{\mathcal{O}}_{p = 0} = \frac{1}{\sqrt{N+1}} \left( \mathcal{O}_{0} + ... + \mathcal{O}_{N} \right).
\end{equation}
which belongs to the same R-symmetry representation as $\mathcal{O}_{\mathrm{pure}}$, namely it is a part of the same protected supermultiplet.

As a side comment, we can now see a further a posteriori justification for our
decoupling constraint $p_{1}+...+p_{I}=0$ on the momentum. Observe that
\textit{if} we had allowed additional excitations in the single impurity
sector, these states would be the highest weight states of a spin $(N-1)/2$
representation of $SU(2)_{\mathcal{R}}$ R-symmetry, and the putative bare
dimension of the operators in these long multiplets would be $2(N+1)$ (see
Appendix \ref{app:superRepresentations}). Observe, however, that in 6D\ SCFTs,
a long multiplet with a scalar of R-charge $J$ has dimension
$4J+6$, so in our case we would be
asserting that these spin $(N-1)/2$ states have dimension greater than $2N+4$,
certainly not a small perturbation to $2N+2$ ! Observe also that no such issue
arises with the spin $(N-3)/2$ representations since in that case the lower
bound for a long multiplet is $2N$, and our operators are well above this
bound. Finally, we note that there is no such gap in scaling dimensions for
4D\ SCFTs, and this is in accord with the fact that imposing a decoupling
limit is not necessary to reach a 4D\ fixed point.

Let us now turn to the case of multiple impurities. Much as in the 4D case, the excitations are characterized by the
Bethe Ansatz equations for the ferromagnetic $XXX_{s = 1/2}$ spin chain. The main distinction is that
now, we need to enforce the condition that the net momentum is zero. Repeating our notation from there, we have the
quasi-particle momenta:%
\begin{equation}
\exp(ip_{j})=\frac{\mu_{j}+\frac{i}{2}}{\mu_{j}-\frac{i}{2}},
\end{equation}
and the 6D decoupling constraint reads as (see subsection \ref{ssec:DECOUP}):
\begin{equation}
\text{6D\ Decoupling Constraint:} \,\,\, \underset{j=1}{\overset{I}{%
{\displaystyle\prod}
}}\left(  \frac{\mu_{j}+\frac{i}{2}}{\mu_{j}-\frac{i}{2}}\right)  =1.
\end{equation}
Other than this, the form of the solutions provided by the Bethe ansatz is the same. Indeed, we still have:
\begin{equation}
\left(  \frac{\mu_{j}+\frac{i}{2}}{\mu_{j}-\frac{i}{2}}\right)  ^{2(N+1)}%
=  \underset{l\neq j}{%
{\displaystyle\prod}
}\frac{\left(  \mu_{j}-\mu_{l}+i\right)  \left(  \mu_{j}+\mu_{l}+i\right)
}{\left(  \mu_{j}-\mu_{l}-i\right)  \left(  \mu_{j}+\mu_{l}-i\right)  },
\end{equation}
for $j=1,...,I$ impurity excitations, and the anomalous dimensions / energy is:
\begin{equation}
(\Delta - \Delta_{0}) = E_{A}=\lambda_{A}\underset{j=1}{\overset{I}{%
{\displaystyle\sum}
}} \epsilon(\mu_j),
\end{equation}
where now $\Delta_0 = 2(N+1)$ and the energy of a given quasi-particle excitation is:
\begin{equation}
\epsilon(\mu) = \frac{i}{\mu + \frac{i}{2}} - \frac{i}{\mu - \frac{i}{2}}.
\end{equation}
We further note that
although $\lambda_{A}$ is quite large, there is a factor of $1/N^2$ for small momenta. This
suppresses the corrections to the anomalous dimensions.
So, at large R-charge this is still a small effect. To get a larger effect one could
of course insert many impurities.

It is also instructive to work out the explicit spectrum of excitations in the special case of
two impurities. Introducing the rapidities $\mu_1$ and $\mu_2$, we note that the 6D decoupling
constraint is readily solved by taking $\mu_1 = -\mu_2 = \mu$. In this case, the Bethe ansatz equations
collapse to a single relation:
\begin{equation}
\left(  \frac{\mu +\frac{i}{2}}{\mu -\frac{i}{2}}\right)  ^{2(N+1)}%
=  - \frac{ 2 \mu + i}{  2 \mu - i } = - \frac{ \mu + \frac{i}{2}}{  \mu - \frac{i}{2} },
\end{equation}
so we learn that the associated momenta are given by:
\begin{equation}\label{eqn:momma}
p_1 = -p_2 = \frac{\pi (2m + 1)}{2N + 1},\,\,\,\text{for}\,\,\, m = 0,...,N.
\end{equation}
We also have the dispersion relation:
\begin{equation}
\epsilon(p) = 4 \sin^{2} \frac{p}{2},
\end{equation}
so in this sector we get anomalous dimensions:
\begin{equation}
(\Delta - \Delta_{0}) = \lambda_{A} \times 8 \sin^{2} \frac{\pi (2m + 1)}{4N + 2}.
\end{equation}
with $m = 0,...,N$.

\subsection{More General Spin Chains}

Much as in the 4D\ case, we can also consider operators which exhibit
additional mixing.\ As one example we can consider operators such as:%
\begin{equation}
	\mathcal{B}_{i}= \sqrt{\mathcal{Z}_{\mathcal{B}_i}} X_{0}....X_{i-1}X_{i}(Y_{i}X_{i})X_{i+1}..X_{N}.
\end{equation}
A similar, though combinatorially more involved analysis follows for mixing in
this case.

We can also include the \textquotedblleft closed-loop\textquotedblright%
\ operators:%
\begin{equation}
	\mathcal{C}_{i,i+L}= \sqrt{\mathcal{Z}_{\mathcal{C}_{i,i+1}}} \text{Tr}(X_{i}....X_{i+L}Y_{i+L}...Y_{i}),
\end{equation}
and we observe a similar local analysis of hopping terms applies. An
interesting feature of the $\mathcal{C}_{i,i+L}$ type operators is that
excitations along the spin
chain should still produce a spectrum with spin chain momentum scaling as
$1/L$. That in turn means that the spectrum of anomalous dimensions will be
controlled by the combination $g_{\text{eff}}^{2}/L^{2}$. By taking $L\sim
N^{1/3+\varepsilon}$, we see that we get a small expansion parameter. As
already mentioned, these sorts of operators have fluctuations which are
visible in the holographic dual.

We note that in both cases, to really trust the analysis in terms of a 1D spin chain, we must suppress possible ``mesonic bubbles''
from forming, as associated with mixing with multi-trace operators. This can be arranged by also assuming the
ranks of the gauge groups are sufficiently large.

\subsection{More General Quivers}

Much as in our discussion of A-type 4D\ SCFTs, we can also consider a more
general class of 6D\ SCFTs in which we vary the ranks of the gauge groups as
we move across the quiver. This leads to a more intricate lattice\ Hamiltonian
since there is a \textquotedblleft middle region\textquotedblright\ where the
ranks are constant, and left and right \textquotedblleft
ramps\textquotedblright\ where the ranks increase. In fact, the set of
possible ramps is in one to one correspondence with nilpotent orbits of the
algebra $\mathfrak{su}(K)$, as in references \cite{Gaiotto:2014lca, DelZotto:2014hpa, Heckman:2016ssk}.
It would be quite interesting to work out the spectrum of the
lattice Hamiltonian in these cases, but we defer this to future work.

\subsection{Little String Theories}

Closely related to our A-type quiver gauge theory is the 6D\ little string
theory (LST)\ obtained by gauging the diagonal subgroup of $G_{0}\times
G_{N+1}$, and introducing an additional non-dynamical tensor multiplet \cite{Bhardwaj:2015oru}.
In this theory, there is an intrinsic string scale as associated with the overall
value of the gauge coupling. This provides a different answer on how to
\textquotedblleft fix the gauge coupling\textquotedblright\ in the
5D\ KK\ regulated theory:\ In some sense it is a free parameter as specified
by the little string theory. Now, in this theory the operator
$\mathcal{O}_{\mathrm{pure}}=\sqrt{\mathcal{Z}_{N}} X_{0}...X_{N}$
is no longer available, but in its place we can construct the
closed loop which winds once around the quiver:
\begin{equation}
	\mathcal{O}_{\text{loop}}= \sqrt{\mathcal{Z}_{N+1}} \text{Tr}(X_{0}...X_{N}).
\end{equation}
We can then work out operator mixing in this theory in much the same way as before. In this
case, we are dealing with a spin chain with periodic boundary conditions. The Bethe ansatz equations
are now given by:
\begin{equation}
\text{6D\ LST Case:} \,\,\, \left(  \frac{\mu_{j}+ \frac{i}{2}}{\mu_{j}- \frac{i}{2}}\right)  ^{N+1}=\underset{l\neq j}{%
{\displaystyle\prod}
}\frac{\mu_{j}-\mu_{l}+i}{\mu_{j}-\mu_{l}-i}.
\end{equation}
And where, as in the case of the 6D SCFT case, we need to take a decoupling limit:
\begin{equation}
p_{1}+...+p_{I}=0,
\end{equation}
which in terms of the rapidities reads as:%
\begin{equation}
\text{6D\ Decoupling Constraint:} \,\,\, \underset{j=1}{\overset{I}{%
{\displaystyle\prod}
}}\left(  \frac{\mu_{j}+ \frac{i}{2}}{\mu_{j}- \frac{i}{2}}\right)  = 1.
\end{equation}

Of course, in the LST case we do not really have a CFT, or even a local
quantum field theory. Nevertheless, at sufficiently low energies we can
characterize the associated effective field theory in terms
of local operators, and our computation reveals that correlation functions
for these local operators are quite similar to those in the closely related 6D SCFT
obtained by decoupling the little string scale. It would be interesting to study this
further.

\section{SCFTs with Conformal Matter \label{sec:CONFTYPE}}

In this section we show that the structure of generalized quivers with
conformal matter points the way to a similar identification of certain
operator subsectors which mix according to a 1D\ spin chain. With this in
mind, we now turn to generalized quivers generated by $N + 1$ M5-branes probing an
ADE\ singularity. On a partial tensor branch where the M5-branes are separated
in the single direction transverse to the singularity, we get a generalized
quiver of the form:%
\begin{equation}
\lbrack G_{0}]-G_{1}-...-G_{N}-[G_{N+1}],
\end{equation}
where here, each gauge group factor corresponds to $G_{i}=G_{ADE}$ for all
$i$, with flavor symmetry factors in the case of $i=0$ and $i=N+1$. We note
that compactifying this theory on a $T^{2}$ results in a 4D\ $\mathcal{N}=2$
SCFT which is also a generalized quiver \cite{Ohmori:2015pua, Ohmori:2015pia, Apruzzi:2018oge}.

In both situations, the geometry of the string realization indicates that
there are operators which still trigger Higgs branch deformations. But as
opposed to the case of theories with A-type matter, in this more general
setting, these operators are not weakly coupled hypermultiplets. This in turn
means that the conformal matter will no longer transform in a doublet
representation of $SU(2)_{\mathcal{R}}$ R-symmetry. In these cases, we instead have
$SU(2)_{\mathcal{R}}$ R-symmetry assignments:%
\begin{equation} \label{tab:spin}
\begin{tabular}
[c]{|c|c|c|c|c|c|}\hline
  & $(A_{K-1},A_{K-1})$ & $(D_{K},D_{K})$ & $(E_{6}, E_6)$ & $(E_7, E_{7})$ &
  $(E_{8}, E_8)$\\\hline
$s$ & $1/2$ & $1$ & $3/2$ & $2$ & $3$\\\hline
\end{tabular}
\ \ \ .
\end{equation}
So in this situation it is fruitful to label each $X_{i}$ as $X_{i}^{(m_{i})}$
where $-s\leq m_{i}\leq s$ denotes the specific spin.
Viewed in this way, we can build a protected highest weight state such as:
\begin{equation}
\mathcal{O}_{\text{pure}}= \sqrt{\mathcal{Z}_N} X_{0}^{(s)}...X_{N}^{(s)},
\end{equation}
but we can also entertain a broad class of impurity insertions. We can label
these according to the $SU(2)_{\mathcal{R}}$ R-symmetry indices as:%
\begin{equation}\label{eqn:6Dop}
	\mathcal{O}_{m_{0},...,m_{N}}= \sqrt{\mathcal{Z}_N} X_{0}^{(m_{0})}...X_{N}^{(m_{N})}.
\end{equation}

We would like to understand operator mixing in a similar fashion to the
quivers with A-type gauge groups. Since we are working to linear order in
perturbations, the structure of these interaction terms are governed by
symmetry considerations. In particular, we expect that the triplet of D-terms
for a given vector multiplet are related to these fields as:%
\begin{equation} \label{eqn:DTERM}
D_{i,a}^{R}=\frac{1}{s}\times\left(  \mathrm{Tr}_{i+1}(X_{i}^{\dag(m_{i})}S_{R}%
^{(m_{i},n_{j})}T_{i,a}X_{i}^{(n_{j})}) - \mathrm{Tr}_{i-1}(X_{i-1}^{(m_{i})}S_{R}%
^{(m_{i},n_{j})}T_{i,a}X_{i-1}^{\dag (n_{j})})\right)  +...
\end{equation}
with $R=1,2,3$ an $SU(2)_{\mathcal{R}}$ R-symmetry triplet index. Here, we have also
included the contributions from the Lie algebra generators. In the above, the
appearance of the \textquotedblleft$...$\textquotedblright\ indicates that we
expect higher order terms due to the fractionation of the M5-branes in the
case of D- and E-type conformal matter.

To extract the structure of hopping terms in this case, we now specialize to the case of a single
impurity insertion, so we focus on operators where all but one of the spins are $s$ and the remaining one has
spin $s-1$. In this case, the calculation is essentially the same as for the A-type quivers,
the only difference is the group theory data associated with bifundamentals of conformal matter
and their associated Goldstone modes.

This is enough to deduce the leading order behavior of the quasi-particle excitations, labelled by momenta $p_{1},...,p_{I}$
for $I$ insertions. We denote by $E_G(\{ p_1 ,..., p_{I} \}) = \Delta - \Delta_0$ the energy associated with this anomalous dimension,
where $\Delta_{0}^{4D} = 2s(N+1)$ and $\Delta_{0}^{6D} = 4s (N+1)$.
We have, in the case of the quiver with $G$-type gauge group:
\begin{equation}
E_{G}(\{ p_1 ,..., p_{I} \}) = \lambda_{G} \underset{1 \leq l \leq I}{\sum} \epsilon(p_i),
\end{equation}
where, for small lattice momentum $p$, we have the approximate
dispersion relation:
\begin{equation}
\epsilon(p) \simeq \frac{p^2}{2s}  + ...,
\end{equation}
where the values of the $\lambda_{G}$ are:
\begin{align}
\lambda_{G}^{4D}  &  = \frac{g_{i}^{2}\widetilde{C}_{G}}{8 \pi^2} \\
\lambda_{G}^{6D}  &  = \frac{g_{\text{eff}}^{2}\widetilde{C}_{G}}{16 \pi^3}  ,
\end{align}
and $g_{\text{eff}}^{2} = (N+1)^{2/3} \pi^{2/3}$.
The relevant values of $\widetilde{C}_G$ are summarized in table \ref{tab:grouptheory}.
\begin{table}%
\centering
\begin{tabular}
[c]{|c|c|c|c|c|c|}\hline
& $SU(K)$ & $SO(2K)$ & $E_{6}$ & $E_{7}$ & $E_{8}$\\\hline
$\dim G$ & $K^{2}-1$ & $K(2K-1)$ & $78$ & $133$ & $248$\\\hline
$d_F$ & $K$ & $K$ & $27$ & $56$ & $248$\\\hline
$f_{G}$ & $1/2$ & $1$ & $3$ & $6$ & $30$\\\hline
$h_{G}^{\vee}$ & $K$ & $2K-2$ & $12$ & $18$ & $30$\\\hline
$\widetilde{C}_{G}$ & $\frac{K^2-1}{2K^2}$ & $\frac{2K-1}{2K-2}$ & $\frac{13}{18}$ & $\frac{19}{24}$ & $1$\\\hline
\end{tabular}
\caption{Relevant group-theoretic quantities appearing in the anomalous dimension.}
\label{tab:grouptheory}
\end{table}

To extract a more precise characterization of operator mixing as well as the associated spin chain
Hamiltonian, we now invoke some special structures present in integrable systems.

\subsection{Spin Chain Hamiltonians}

Thus far we have presented evidence that the one loop corrections to the
anomalous dimensions of the large R-charge operators of line
(\ref{eqn:6Dop}) can be understood via a 1D open spin chain with nearest-neighbour interactions.
We have studied the operators in the A-type conformal matter theory in
detail, and the main difference between the D- and E-type theories is that
we expect, on general grounds, that there could be additional spin excitations
which contribute to operator mixing.

For the A-type conformal matter theories we have shown that this spin chain
consists of $\ket{\uparrow}$ and $\ket{\downarrow}$ spin states at each site
and has a Hamiltonian:
\begin{equation}
  H_{A}= - \lambda_{A} \underset{i=0}{\overset{N-1}{\sum}%
} \left( 2 \overrightarrow{S}_{i}\cdot\overrightarrow{S}_{i+1} - \frac{1}{2} \right) \,.
\end{equation}
As we have already remarked, this is the Hamiltonian of the ferromagnetic $XXX_{1/2}$
Heisenberg spin chain with open boundary conditions, and it is well known that this system is, in fact
integrable!

From the perspective of the holographic duals defined by $AdS_7 \times S^4 / \Gamma$,
there is not much difference between excitations passing from the north pole to the south pole
in the cases of the different orbifold groups. So, we will make the reasonable assumption that the spin chain relevant
for the $(D_K, D_K)$ conformal matter operators is also integrable. Indeed,
once this assumption is made the contributions from the higher order
interactions to the Hamiltonian are fixed. In this case we are dealing with
an open spin chain consisting of $N+1$ sites each of which hosts a spin in the
$s=1$ representation of $SU(2)_{\mathcal{R}}$, and which has only nearest-neighbour
interactions. Assuming we have an integrable system, this is nothing but the $XXX_s$ spin
chain, and one of the triumphs of the algebraic Bethe ansatz is that one can
uniquely determine the form of the Hamiltonian of such a system. With our
present conventions it is given by (see e.g. the review \cite{Faddeev:1996iy}):
\begin{equation}
H_{D}= - \lambda_{D}\underset{i=0}{\overset{N-1}{\sum}%
}\left( \frac{1}{2} \overrightarrow{S}_{i}\cdot\overrightarrow{S}_{i+1}- \frac{1}{2} \left(
\overrightarrow{S}_{i}\cdot\overrightarrow{S}_{i+1}\right)  ^{2}\right)
\label{HDOP}%
\end{equation}
where here, the $\overrightarrow{S}_{i}$ describe spin $s=1$
excitations. The relative coefficients of these two terms are fixed by
the condition of integrability. The overall normalization of the coupling is fixed by the demand that
we get the correct dispersion relation. From our normalization of the generalized D-term potential given in
equation (\ref{eqn:DTERM}), we expect $\epsilon(p) \simeq p^{2} / 2s$, with $s = 1$ in the case of $G = SO(2K)$.
As we show later in subsection \ref{ssec:BETHE}, this is in accord with the relation which links the
rapidities of the Bethe ansatz to energies of quasi-particle
excitations:
\begin{equation}
\epsilon(\mu) = \frac{d}{d \mu} p(\mu).
\end{equation}

The precise form of the dilatation operator would be hard to guess a priori, but
we can motivate the appearance of such a term, at least from the perspective
of conformal matter for D-type theories. Observe that on the full tensor
branch, the D-type quivers (For $SO(2K)$ with $K>4$) consist of alternating
$SO/Sp$ gauge group factors of the form:%
\begin{equation}
\lbrack SO]-Sp-SO-...-SO-Sp-[SO].
\end{equation}
Between each such gauge group factor we have weakly coupled half
hypermultiplets in the bifundamental representation. Viewing each such
bifundamental as a spin $s=1/2$ excitation, the composite operator obtained
from a product of two such operators transforms in the $s=1$ or $s=0$
representation. Now, given this, we might attempt to analyze our system in
terms of an $XXX_{s=1/2}$ spin chain of double the length. If we now perform a
block spin decimation procedure we can instead attempt to work in terms of the
composite $s=1$ excitations. Doing so, higher order terms become somewhat
inevitable, and the precise form demanded by integrability is that of line
(\ref{HDOP}).

Having come this far, it is now just a further small jump to demand the same
structure also persists in the case of the E-type theories. Indeed, from the
perspective of $AdS_{7}\times S^{4}/\Gamma_{ADE}$ we expect little difference
in our protected subsector, especially between the D- and E-type cases. With
this in mind, we now simply assume that the other cases are also governed by an integrable
$XXX_{s}$ spin chain. Figuring out the dilatation operator responsible for operator
mixing then means determining the corresponding integrable spin chain
Hamiltonian. The end result was obtained using the algebraic Bethe ansatz
in \cite{Babujian:1982ib} (see also the review \cite{Faddeev:1996iy},
modulo a few unfortunate typos\footnote{We thank V. Korepin for helpful
comments.}), and we will take our answer from there.

The integrable $XXX_s$ spin chain for $s \geq 1$ has been studied
in great detail (see references \cite{Zamolodchikov:1980ku, Kulish:1981gi, Kulish:1981bi, Babujian:1982ib, A.:1982zz, Babujian:1983ae, Tarasov:1983cj}). The integrable $XXX_{s}$ spin chain Hamiltonian takes the form:
\begin{equation}
H_{G}=-\lambda_G \sum_{n=0}^{N-1} Q_{2s}(\overrightarrow{S}_{n}\cdot
\overrightarrow{S}_{n+1})\,,
\end{equation}
where $G$ refers to our choice of gauge group, which is linked to a choice of spin $s$ (as already indicated above)
and we remind the reader that we are labelling the $(N+1)$ sites from $0$ to
$N$. The $S_{n}=(S_{n}^{x},S_{n}^{y},S_{n}^{z})$ are the spin $s$ operators at
the $n^{th}$ site, and $Q_{2s}$ is a degree $2s$ polynomial. The overall normalization
by the pre-factor $\lambda_{G}$ has been chosen so that we again retain the quasi-particle
dispersion relation $\epsilon(p) = p^{2} / 2s$ which is in accord with the expression:
\begin{equation}
\epsilon(\mu) = \frac{d}{d \mu} p(\mu),
\end{equation}
with $\mu$ a Bethe ansatz rapidity (see subsection \ref{ssec:BETHE}).

The polynomial $Q_{2s}$ is chosen such that the energy of the ferromagnetic
ground state vanishes, that is,
\begin{equation}
Q_{2s}(s^{2})=0\,.
\end{equation}
The structure of the spin chain Hamiltonian is then fixed by demanding
integrability. As reviewed in \cite{Faddeev:1996iy} (our
presentation follows reference \cite{Korepin:2004zz}):
\begin{equation}
Q_{2s}(x)= - 2\underset{l = 0 }{\overset{2s}{\sum}}%
\underset{k= l +1}{\overset{2s}{\sum}}\frac{1}{k}\underset{j\neq
l}{\underset{j=0}{\overset{2s}{%
{\displaystyle\prod}
}}}\frac{x-x_{j}}{x_{l}-x_{j}},\text{ \ \ with \  }x_{l}=\frac{1}%
{2}l(l+1)-s(s+1).
\end{equation}

While reviewing the method of finding this formula would take us too far
afield (see e.g. \cite{Faddeev:1996iy}), we simply note that the appearance of sums and products up to $2s$ has
to do with taking irreducible representations from the Clebsch-Gordon
decomposition $s\otimes s=2s\oplus...\oplus0$.

Plugging in for the various cases of interest to us and using the correspondence between different
spin assignments $s$ and the corresponding ADE gauge group (see line \ref{tab:spin}), we get:%
\begin{align}
Q_{A_{k}}(x) &  =-\frac{1}{2}+2x\\
Q_{D_{k}}(x) &  =+\frac{1}{2}x-\frac{1}{2}x^{2}\\
Q_{E_{6}}(x) &  =-\frac{3}{4}-\frac{1}{8}x+\frac{1}{27}x^{2}+\frac{2}{27}%
x^{3}\\
Q_{E_{7}}(x) &  =-\frac{1}{2}+\frac{13}{24}x+\frac{43}{432}x^{2}-\frac{5}%
{216}x^{3}-\frac{1}{144}x^{4}\\
Q_{E_{8}}(x) &  =-\frac{148}{125}-\frac{1687}{9000}x+\frac{1297}{18000}%
x^{2}+\frac{593}{20250}x^{3}+\frac{79}{97200}x^{4}-\frac{77}{243000}%
x^{5}-\frac{1}{48600}x^{6},
\end{align}
in the obvious notation. From this, we obtain the nearest neighbor spin chain
Hamiltonian in all cases.

The appearance of higher order spin-spin interaction terms is quite
non-trivial but is again in accord with expectations where we view conformal
matter excitations as a ``composite object'' built out of small spin excitations.

Having fixed the form of our spin chain Hamiltonian in the uniform case, we
can also conjecture that there is a natural generalization of these
considerations in which we allow a position dependent coupling in the spin
chain. This structure is expected in the various generalized quiver theories.
In this setting, we have one final generalization:%
\begin{equation}
H= -\underset{i}{\sum}\lambda_{i}H_{i,i+1} \,,
\end{equation}
where $H_{i,i+1}$ encodes all the nearest-neighbour interactions between the
subscripted sites, and we are now allowing position dependent couplings along
the ``ramps'' of the generalized quiver. On the plateau with all equal ranks,
however, these couplings are independent of the spin site.

\subsection{Bethe Ansatz \label{ssec:BETHE}}

The passage to the open $XXX_{s}$ spin chain follows the same path
already discussed for the $XXX_{s = 1/2}$ case, and also
follows the presentation given in \cite{Faddeev:1996iy}. The relevant definitions in this case
relating the quasi-particle momenta and rapidities are:%
\begin{equation}
\exp(ip_{j})=\frac{\mu_{j}+is}{\mu_{j}-is}.
\end{equation}
With conventions as before, the energy / anomalous dimension of an
excitation is now given by:
\begin{equation}
(\Delta - \Delta_0) = E_{G}=\lambda_{G}\underset{j=1}{\overset{I}{%
{\displaystyle\sum}
}}\left(  \frac{i}{\mu_{j}+is}-\frac{i}{\mu_{j}-is}\right)  ,
\end{equation}
where $\Delta_{0} = 4s(N+1)$,
and the specific values of the rapidities are fixed by the Bethe ansatz
equations:%
\begin{equation}
\left(  \frac{\mu_{j}+is}{\mu_{j}-is}\right)  ^{2(N+1)}= \underset{l\neq j}{%
{\displaystyle\prod}
}\frac{\left(  \mu_{j}-\mu_{l}+i\right)  \left(  \mu_{j}+\mu_{l}+i\right)
}{\left(  \mu_{j}-\mu_{l}-i\right)  \left(  \mu_{j}+\mu_{l}-i\right)  }.
\end{equation}

As before, we can also consider the case of periodic boundary conditions, corresponding
to a little string theory. In that case, the
relation between energies and rapidities is unchanged, but the rapidities now
satisfy the equation:%
\begin{equation}
\text{6D\ LST Case:} \,\,\,\left(  \frac{\mu_{j}+is}{\mu_{j}-is}\right)  ^{N+1}=\underset{l\neq j}{%
{\displaystyle\prod}
}\frac{\mu_{j}-\mu_{l}+i}{\mu_{j}-\mu_{l}-i}.
\end{equation}
Again, we remark that in the LST case we do not have a genuine local quantum field theory, but the correlation
functions of the low energy effective field theory are nevertheless well captured by the same
structure as their closely related 6D SCFT cousins.

So far, we have focused on some general features of the $XXX_{s}$ spin chain
and its spectrum of excitations. Now, in the specific application to 6D\ SCFTs
as well as 6D\ LSTs, we need to impose a further constraint to achieve a
proper decoupling limit, namely, we need to set (see subsection \ref{ssec:DECOUP}):%
\begin{equation}
\text{6D\ Decoupling Constraint:} \,\,\, p_{1}+...+p_{I}=0,
\end{equation}
which in terms of the rapidities reads as:%
\begin{equation}
\text{6D\ Decoupling Constraint:} \,\,\, \underset{j=1}{\overset{I}{%
{\displaystyle\prod}
}}\left(  \frac{\mu_{j}+is}{\mu_{j}-is}\right)  =1.
\end{equation}
We again note that in the 4D\ theories we do not impose this constraint since we
have a marginal coupling available which allows us to tune the operator
spectrum close to the free field limit.

\subsection{Two Impurity Sector for 6D SCFTs}

To close this section, we consider in more detail the case of 6D\ SCFTs with
two impurity insertions, much as we did in section \ref{sec:6DATYPE}.
We observe that just as in the A-type quiver gauge theories, our
system of equations collapses to a single condition under the assumption
$\mu = \mu_{1}=-\mu_{2}$. The remaining Bethe ansatz equation is then given by:%
\begin{equation}\label{ImpishDimpishDurpyDoo}
\left(  \frac{\mu+is}{\mu-is}\right)  ^{2(N+1)}=-\frac{\mu+\frac{i}{2}}%
{\mu-\frac{i}{2}}.
\end{equation}
The case of $s > 1/2$ is somewhat more challenging to solve than the $s = 1/2$ case
considered in equation (\ref{eqn:momma}). To
proceed further, it is convenient to perform a formal expansion in powers of
$1/N$ as well as $1/s$. Doing so, we find that the momentum $p=p_{1}=-p_{2}$ satisfies:\footnote{To arrive at this expression, 
write $e^{ip} = (\mu + is)/(\mu - is)$. Solving for $\mu$ yields $\mu = s \cot (p/2) $. 
Then, substituting into the righthand side of equation (\ref{ImpishDimpishDurpyDoo}), 
we can systematically solve order by order in a $1/s$ expansion.}
\begin{equation}
p=\frac{\pi (2m + 1)}{2N+2}\left(  1+ \frac{1}{2s}\frac{1}{2N+2}+...\right)  ,
\end{equation}
where $m=0,...,N$. The energy in the two impurity sector is then given by:%
\begin{equation}
\left(  \Delta-\Delta_{0}\right)  =E_{G}=\lambda_{G}\times\frac{4}{s}\sin
^{2}\frac{p}{2},
\end{equation}
where $\Delta_{0}=4s(N+1)$, and we remind the reader that
\begin{equation}
\lambda_{G} = \frac{(N+1)^{2/3}\pi^{2/3} \widetilde{C}_{G}}{16 \pi^3} .
\end{equation}

\newpage

\section{Conclusions \label{sec:CONC}}

The study of 6D\ SCFTs has led to great progress in the understanding of
quantum field theory. This is all the more remarkable considering that the
only known realizations of such theories rely on string theory. In
this paper we have used the generalized quiver description of 6D\ SCFTs to
extract some information on the operator content of these theories. We have
also applied a similar set of tools in the case of 4D\ SCFTs obtained from
compactification of their partial tensor branch deformations on a $T^{2}$. In particular, we have argued
for the existence of nearl protected operator subsectors at large R-charge which have scaling dimensions
controlled by a perturbation series in inverse powers of the R-charge.
Introducing a 5D\ KK\ regulator for 6D SCFTs, we have shown how to extract a
corresponding operator mixing matrix. In the case of quivers with A-type gauge
groups, we have shown that this leads to a mixing matrix which is
mathematically identical to the $XXX_{s=1/2}$ Heisenberg spin chain with open
boundary conditions. Moreover, by appealing to the similar structure present
in generalized quivers with D- and E-type gauge groups we have extended our
considerations to these cases as well. Assuming the existence of the
corresponding integrable structures, we have shown how to extract the operator
scaling dimensions for certain subsectors of our 6D\ SCFTs. In the remainder
of this section we discuss some avenues of future investigation.

In this paper we have mainly focused on the spectrum of excitations above the
\textquotedblleft ground state\textquotedblright\ operator given by
$\mathcal{O}_{\text{pure}}= \sqrt{\mathcal{Z}_{N}} X_{0}...X_{N}$. We have also seen that similar
operator mixing effects exist for other protected and \textquotedblleft nearly
protected\textquotedblright\ operators. It would be very interesting to
extract the spectrum of anomalous dimensions for these cases as well.

Focusing on the case of quivers with A-type gauge groups, we have also
observed that there is another natural class of spin chain Hamiltonians which
we obtain by allowing position dependent coupling constants. This can occur
because such quivers can have long \textquotedblleft ramps\textquotedblright%
\ in which the ranks of gauge groups slowly increase as we reach the interior
region of a long quiver. Developing a suitable extension of the Bethe ansatz
in such situations is an immediate goal in this direction.

One of the general messages of this work is that the appearance of spin chains
in these systems is in close accord with the quiver-like structure of these
theories. Given this, it is tempting to consider Higgs branch flows which
connect these theories to the $\mathcal{N}=(2,0)$ SCFTs, and in so doing,
extract additional details on the resulting operator content.

There is a natural generalization of our results to a broader
class of spin excitations. Treating the operator $\mathcal{O}_{\text{pure}%
}=\sqrt{\mathcal{Z}_{N}}X_{0}...X_{N}$ as the ground state for our spin chain,
we can ask about the effects of adding more general sorts of impurity
excitations, as captured by the complexification of the superconformal algebra
$\mathfrak{osp}(8^{\ast}|1)$. The related question has already been discussed
in the context of $\mathcal{N}=4$ super Yang--Mills in reference
\cite{Minahan:2002ve, Beisert:2003yb}, and we can adapt these considerations
to the present case. The more general sorts of impurities correspond to
swapping out an $X_{i}$ for a covariant derivative insertion such as $D_{\mu
}X_{i}$, as obtained from an excitation in the $\mathfrak{so}(6,2)\subset
\mathfrak{osp}(8^{\ast}|1)$ subalgebra. Additionally, we can insert fermionic
states which are associated to the $\mathbb{Z}_{2}$ odd part of the superalgebra.
In fact, the relevant spin chain analysis for this superalgebra has been
carried out both for periodic \cite{Martins:1997wb} and open boundary
conditions \cite{Arnaudon:2003zw}. This will likely provide a point of entry
for accessing more precise information on the spectrum of operators in 6D
SCFTs. There are other natural operator sectors which appear amenable to a spin chain
analysis. It would be interesting to also cast these cases in the language of
integrable super spin chains.

We have mainly focused on the leading order effects in perturbation
theory, but one can also entertain extending this analysis to higher
orders in perturbation theory, as associated with next to nearest neighbor
interactions. At least in the limit of large R-charge, we again have a
perturbative expansion parameter, so we can in principle contemplate the form
such operator mixing effects must take.

From the perspective of top down constructions, the 6D SCFTs considered here
are all realized as edge modes in a higher-dimensional system. Lifting our
discussion of integrability to this setting suggests a potential way of
arguing from first principles for the appearance of such integrable
structures. It would be interesting to develop this perspective and explore
potential connections to recent higher-dimensional perspectives on
integrability such as references \cite{Costello:2013sla, Costello:2013zra,
Costello:2017dso, Costello:2018gyb}.

The appearance of integrable 1D\ spin chains in 6D\ SCFTs is by itself quite
intriguing. It is also natural to consider possible deformations of such
integrable systems, and their relation to deformations of 6D SCFTs. This would
likely lead to an improved understanding of more general phenomena associated with
quantum fields in diverse spacetime dimensions.

\section*{Acknowledgements}

We thank V. Korepin for helpful correspondence. We thank M.J. Kang for comments
on an earlier draft. The work of FB is supported by
the Spanish Research Agency (Agencia Estatal de Investigaci\'{o}n) through the
grant IFT Centro de Excelencia Severo Ochoa SEV-2016-0597, and by the grant
PGC2018-095976-B-C21 from MCIU/AEI/FEDER, UE. The work of JJH and CL is
supported by a University Research Foundation grant at the University of Pennsylvania.

\newpage

\appendix

\section{6D Superconformal Unitary Representations}

\label{app:superRepresentations}

The six-dimensional $\mathcal{N}=(1,0)$ superconformal algebra is given by
$\mathfrak{osp}(8^{\ast}|1)$. The bosonic subalgebra is $\mathfrak{so}%
(6,2)\oplus\mathfrak{sp}(1)_{R}$. By convention, we choose \emph{half}%
-integers, $J \in\mathbb{N}/2$, to label R-symmetry representations of
$\mathfrak{sp}(1)_{R}\simeq\mathfrak{su}(2)$, and integer valued Dynkin labels,
$[j_{1},j_{2},j_{3}]$, for representations of the Lorentz group. We refer to
\cite{Cordova:2016emh, Buican:2016hpb} and references therein for more details
on the construction of unitary representations of the superconformal algebras
in various dimensions.

A generic six-dimensional superconformal multiplet is denoted by:
\begin{equation}
\chi\lbrack j_{1},j_{2},j_{3}]_{\Delta}^{J}\,,
\end{equation}
with $\Delta$ the dimension of the superconformal primary. Unitarity imposes
restrictions on the possible values of the dimension of a multiplet. For a
long multiplet, $\mathcal{L}$, it imposes a bound from below:
\begin{equation}
\mathcal{L}[j_{1},j_{2},j_{3}]_{\Delta}^{J}:\qquad\Delta > 4J+\frac{1}%
{2}(j_{1}+2j_{2}+3j_{3})+6\,,
\end{equation}
In addition to long multiplets there exist short multiplets with conformal
dimensions set by the R-symmetry and Lorentz quantum numbers. The simplest
are $\mathcal{A}$-type multiplets, short multiplets at threshold:
\begin{equation}
\mathcal{A}[j_{1},j_{2},j_{3}]_{\Delta}^{J}:\qquad\Delta=4J+\frac{1}{2}%
(j_{1}+2j_{2}+3j_{3})+6\,,
\end{equation}
In six dimensions there are then three additional isolated short multiplets.
In these cases there exist superconformal descendants that are annihilated by
specific combinations of the supercharges, allowing for a conformal dimension
below that of a long multiplet:
\begin{align}
\mathcal{B}[j_{1},j_{2},0]_{\Delta}^{J}: &  \qquad\Delta=4J+\frac{1}{2}%
(j_{1}+2j_{2})+4\,,\\
\mathcal{C}[j_{1},0,0]_{\Delta}^{J}: &  \qquad\Delta=4J+\frac{1}{2}%
j_{1}+2\,,\\
\mathcal{D}[0,0,0]_{\Delta}^{J}: &  \qquad\Delta=4J\,.
\end{align}
The superconformal primary of $\mathcal{D}$-type multiplets is annihilated by
half of the supercharges and therefore is 1/2-BPS.

\newpage

\section{One Loop Diagram \label{app:LOOP}}

In this Appendix we evaluate:%
\begin{equation}
I(x)=\int d^{D}z\text{ }\frac{|x|^{4\Delta}}{|x-z|^{4\Delta}|z|^{4\Delta}}.
\end{equation}
Working in Euclidean signature, we can write:%
\begin{equation}
I(x)=\underset{\text{Min}}{\overset{\text{Max}}{\int}}\left\vert z\right\vert
^{D-1}d\left\vert z\right\vert \underset{0}{\overset{\pi}{\int}}\sin
^{D-2}\theta d\theta\underset{S^{D-2}}{\int}d\Omega_{D-2}\text{ }%
\frac{|x|^{4\Delta}}{\left| z^{2}+x^{2}-2|z| |x|\cos\theta\right| ^{2\Delta
}\,|z|^{4\Delta}}.
\end{equation}
We observe that when $D = 4$ and $\Delta = 1$ as well as when $D = 6$ and $\Delta = 3/2$
there is a logarithmic divergence, as obtained by evaluating the integral near the UV and IR cutoffs.
Since we are only interested in the logarithmic divergence anyway, we are left with the
integral (see e.g. \cite{Berenstein:2002jq, Minahan:2002ve}):
\begin{equation}
I(x) \approx \underset{\Lambda^{-1}}{\overset{\vert x \vert}{\int}} \frac{d \xi}{\xi} \underset{S^{D-1}}{\int} d \Omega_{D-1}.
\end{equation}
Evaluating in the two cases of interest, we have:
\begin{align}
D=4:I(x) & \approx (\Omega_{3})\times\log(\left\vert x\right\vert \Lambda) + ...,\\
D=6:I(x) & \approx (\Omega_{5})\times\log(\left\vert x\right\vert \Lambda) + ...,
\end{align}
where $\Omega_{3} = 2 \pi^2$ denotes the volume of a unit radius $S^{3}$ and $\Omega_{5} = \pi^{3}$ denotes the
volume of a unit radius $S^5$.

\newpage

\bibliographystyle{utphys}
\bibliography{6DSpinChains}

\end{document}